\documentclass[pdflatex,sn-vancouver-num]{sn-jnl}

\usepackage{graphicx}%
\usepackage{multirow}%
\usepackage{amsmath,amssymb,amsfonts,amsthm}%
\usepackage{mathrsfs}%
\usepackage{graphicx}
\usepackage{bookmark,booktabs}
\usepackage{manyfoot}%
\usepackage{xcolor}
\usepackage{tabularx}
\usepackage{hyperref}
\usepackage{subcaption}
\usepackage[normalem]{ulem}

\theoremstyle{thmstyleone}%
\theoremstyle{thmstyletwo}%

\theoremstyle{thmstylethree}%

\begin{document}

\title[Modeling Physical Activity Change as Smooth Transformations]{Modeling Physical Activity Change as Smooth Transformations: Temporal and Amplitude Patterns Associated with Physical Function in Older Women}

\author[1]{\fnm{Rong W.} \sur{Zablocki}}

\author[1]{\fnm{Steve} \sur{Nguyen}}

\author[1]{\fnm{Yacun \sur{Wang}}}

\author[1]{\fnm{Lindsay} \sur{Dillon}}

\author[2]{\fnm{Michael J.} \sur{LaMonte}}

\author[3]{\fnm{Phyllis A.} \sur{Richey}}

\author[4]{\fnm{Ramon} \sur{Casanova}}

\author[5]{\fnm{Marcia L.} \sur{Stefanick}}

\author[1]{\fnm{Sheri J.} \sur{Hartman}}

\author[6]{\fnm{Chongzhi} \sur{Di}}

\author[6]{\fnm{Charles} \sur{Kooperberg}}

\author[1]{\fnm{Loki \sur{Natarajan}}}

\author[1]{\fnm{Andrea Z.} \sur{LaCroix}}

\author*[1]{\fnm{Jingjing} \sur{Zou}\email{j2zou@ucsd.edu}}

\affil[1]{\orgdiv{Herbert Wertheim School of Public Health and Human Longevity Science}, \orgname{University of California at San Diego}}

\affil[2]{\orgdiv{Department of Epidemiology and Environmental Health}, \orgname{State University of New York at Buffalo}}

\affil[3]{\orgdiv{Department of Preventive Medicine}, \orgname{The University of Tennessee Health Science Center}}

\affil[4]{\orgdiv{Biostatistics and Data Science, Public Health Sciences}, \orgname{Wake Forest University School of Medicine}}

\affil[5]{\orgdiv{Stanford Prevention Research Center}, \orgname{Stanford University School of Medicine}}

\affil[6]{\orgdiv{Division of Public Health Sciences}, \orgname{Fred Hutchinson Cancer Center}}

\abstract{\textbf{Purpose:} To investigate whether longitudinal changes in timing and magnitude of PA are associated with physical function (PF) in older women.

\textbf{Methods:} Women from Objective Physical Activity and Cardiovascular Health (OPACH) study with accelerometry at baseline and Women’s Health Initiative Strong and Healthy (WHISH) W1 and W2 were included. Minute-level PA counts were averaged and smoothed as diurnal PA curves. Consecutive-visit change was modeled within periods (baseline--W1 and W1--W2) as a Riemannian deformation from earlier to later curves, with two-dimensional initial momenta characterizing timing and magnitude shifts. Multivariate functional principal component analysis (MFPCA) summarized coupled timing-magnitude patterns, and principal component (PC) scores and deformation energy were derived for each participant-period. Linear mixed-effects models related these features to RAND-36 PF, adjusting for baseline PF and covariates.

\textbf{Results:} Mean PA deformation in both periods showed downward shifts in PA magnitude and temporal redistribution after 10:00. Top 15 PCs explained at least 90\% of variability in both periods. PC1 captured diurnal PA increase/decrease, explaining 22.4\% of variability for baseline--W1 and 20.8\% for W1--W2. 
Among participants with complete PF scores and baseline covariates ($N=1{,}157$), higher PC1 scores, reflecting relative increase/maintenance of PA across day, were positively associated with PF ($P<0.0001$). Deformation energy, a metric for overall diurnal pattern change between visits, showed a significant interaction with period for PF ($P=0.003$), with a larger positive association during W1--W2 than during baseline--W1. 

\textbf{Conclusions:} In older women, longitudinal changes in diurnal PA accumulation were associated with PF. Riemannian deformation analysis identified clinically interpretable markers of PA pattern change that may capture functional-aging information not represented by conventional PA summaries.}


\maketitle

\clearpage
\section{Background}\label{intro}

Regular physical activity (PA) is a central behavioral target for promoting cardiovascular health, preserving function, and extending healthy aging \citep{warburton2017health, bull2020world}. Conversely, physical inactivity and sedentary behavior (SB) are associated with higher risks of cardiovascular disease (CVD), all-cause mortality, cancer, obesity, and declines in physical and cognitive function in older adults \citep{LaMonte:2018jb, LaCroix:2019eb, Chastin:2019bw, nguyen2024prospective, lamonte2024accelerometer, hyde2025sitting, silveira2022sedentary, saunders2020sedentary, raichlen2025associations}. 
Physical function (PF) is especially relevant in aging research because it reflects the capacity to perform daily activities and is closely linked to mortality and risks of diseases such as CVD \citep{kritchevsky2019pathways}. Characterizing how free-living PA changes over time may therefore help identify clinically meaningful patterns particularly in older adults who experience heterogeneous trajectories of functional decline and cardiometabolic risk \citep{leskinen2022daily, hoekstra2020distinct, wu2021cardiometabolic}.

Accelerometers provide directly measured, high-resolution PA data in free-living settings. However, common analytic approaches often reduce minute-level time series to daily summaries, such as total activity volume, minutes per day above intensity cut-points, or time spent sedentary \citep{fuezeki2017health, evenson2015calibrating, karas2022comparison, lamonte2024accelerometer}. These measures are interpretable, but they may obscure diurnal patterns of PA. Two individuals with similar total PA may differ substantially in when activity occurs, whether activity is maintained across the day, and how activity routines shift with aging. Such timing and magnitude changes may be important in older women, for whom functional decline may manifest not only as lower total activity, but also as altered activity patterns.

Functional data analysis provides a natural framework for studying accelerometer profiles as continuous diurnal curves rather than summary metrics. Functional principal component analysis (FPCA) has been used to identify major modes of variation in minute-level PA profiles \citep{lin2022longitudinal, zablocki2024using, acar2025functional}. Standard FPCA, however, typically decomposes PA curves across subjects and visits jointly, so longitudinal change is inferred indirectly from differences in participant-level scores rather than examined directly.
In addition, without explicit alignment or time-warping of curves, standard FPCA does not separate changes in PA timing, or temporal variation along the $x$-domain, from changes in PA magnitude, or amplitude variation along the $y$-domain. A framework that models the visit-to-visit transformation directly may better capture clinically interpretable changes in daily activity organization.

Riemannian deformation offers such an approach by representing each smoothed diurnal PA profile as a geometric object and modeling change between visits as a smooth transformation from an earlier curve to a later curve \citep{zou2023riemann, beg2005computing, charlier2017fshape}. This formulation preserves within-day temporal structure while jointly characterizing horizontal shifts in activity timing and vertical shifts in activity magnitude. It is therefore well suited for aging cohorts, in which decline, compensation, and routine adaptation may appear as coordinated changes in both \emph{when} and \emph{how much} PA is accumulated \citep{erickson2024age, luo2025aging}. Cohort-level heterogeneity in these transformations can then be summarized through principal component analysis of the deformation features.

In this study, we use the repeated free-living accelerometer and PF assessments in Objective Physical Activity and Cardiovascular Health  and Women's Health Initiative Strong and Healthy(OPACH-WHISH) studies to examine visit-to-visit PA change in older women. We extend the Riemannian deformation framework of \citet{zou2023riemann} by integrating multivariate FPCA (MFPCA) \citep{happ2018multivariate}, allowing temporal and amplitude components of PA change to be modeled jointly as a multivariate functional object. We treated PA change from baseline to W1 and from W1 to W2 as the primary longitudinal phenotype, with the objectives of identifying main patterns of change in diurnal PA timing and magnitude and determining whether participant-level deformation features were associated with PF. This OPACH-WHISH centered analysis aims to translate repeated accelerometer measurements into interpretable markers relevant to risk stratification and PA intervention targeting in older women.

\section{Methods}\label{sec:methods}
\subsection{Study design and participants}

The OPACH study was a Women's Health Initiative (WHI, initiated in 1992 \citep{study1998design}) ancillary study that enrolled 7,048 ambulatory, community-dwelling women aged 63–99 years between 2012 and 2014 \citep{lacroix2017objective, LaMonte:2018jb, nguyen2024prospective}. Among surviving OPACH participants, 
2,356 consented to participate in the accelerometer sub-study of the WHISH trial (initially funded from 2015-2020) \citep{stefanick2021women, stefanick2022women, wegner2021physical}, with accelerometer deployments scheduled at 6, 18, and 36 months post-randomization. The focus of WHISH was to test whether increasing PA reduced the risk of primary CVD events in older women. Together with the OPACH baseline assessment, this design afforded up to four waves of accelerometer-measured PA data per participant. All participants provided written informed consent, and the studies were approved by the Institutional Review Board at the Fred Hutchinson Cancer Center \citep{lacroix2017objective, stefanick2022women}.

The present analyses focus on the first three accelerometer assessment waves: OPACH baseline (original $n_0$ = 6,489), WHISH wave 1 at 6 months post-randomization (W1; $n_1$ = 1,934), and WHISH wave 2 at 18 months post-randomization (W2; $n_2$ = 1,677). The average interval between baseline to W1 was 3 years and the average between W1 to W2 was 1 year. We refer to this analytic sample as the OPACH–WHISH sub-cohort. Data from the WHISH wave 3 at 36-month (W3; $n_3$ = 1,398) were excluded due to substantial loss to follow-up, to minimize potential bias in FPCA estimation and subsequent association analyses. At the time of the present analysis, WHISH randomization assignments (usual activity versus PA intervention) for OPACH–WHISH sub-cohort members had not yet been made available to the study team; implications are addressed in the Conclusion (\ref{sec:discussion}).

\subsection{Accelerometer data and pre-processing} \label{qc_process}
Participants were provided with an ActiGraph GT3X+ (Pensacola, Florida) tri-axial accelerometer and instructed to wear it over the right hip, 24 hours per day for 7 days at each assessment wave, only to remove it for bathing and other water-based activities \citep{lacroix2017objective, stefanick2021women}. The device captures acceleration along three orthogonal axes: vertical, horizontal, and perpendicular (VA, HA, PPA) \citep{kuster2020sitting, migueles2017accelerometer}. 15-second epoch count files were generated via ActiLife software (v.6) \citep{lacroix2017objective, stefanick2021women}. Counts from each axis were summed over four consecutive 15-second intervals to obtain minute-level counts, then combined across axes to compute the vector magnitude (VM) as $\sqrt{\mbox{VA}^2+\mbox{HA}^2+\mbox{PPA}^2}$ \citep{sasaki2011validation}. Non-wear windows were identified using the Choi algorithm \citep{choi2011validation, choi2012assessment}.

\textbf{Inclusion criteria and valid-day definitions.} For each wave, minute-level counts from 6:00 a.m. to midnight (18 hours) were retained, as activity outside this window predominantly reflected sleep or in-bed time \citep{zou2023riemann}. Days with $>$4 hours of non-wear between 6:00 a.m. and midnight were excluded (corresponding to the $90^{\mathrm{th}}$ percentile of daily non-wear), and a valid day required $\ge$14 hours of wear time within this window. Participants were required to have at least four valid days per visit to be included \citep{migueles2017accelerometer,LaCroix:2019eb}. These criteria yielded $N_1=1{,}646$ participants with valid data at both baseline and W1, and $N_2=1{,}362$ with valid data at both W1 and W2 (84 of whom lacked qualified baseline data). The final accelerometer analytic sample included $N^+=1{,}730$ $(1,646 + 84)$ participants, each with at least two consecutive non-missing visits (mean valid days per visit: 6.3; median: 7; range: 4--10). 

\begin{figure}[h]
  \begin{subfigure}[b]{0.5\textwidth}
    \includegraphics[width=\linewidth,height=0.41\textheight, keepaspectratio]{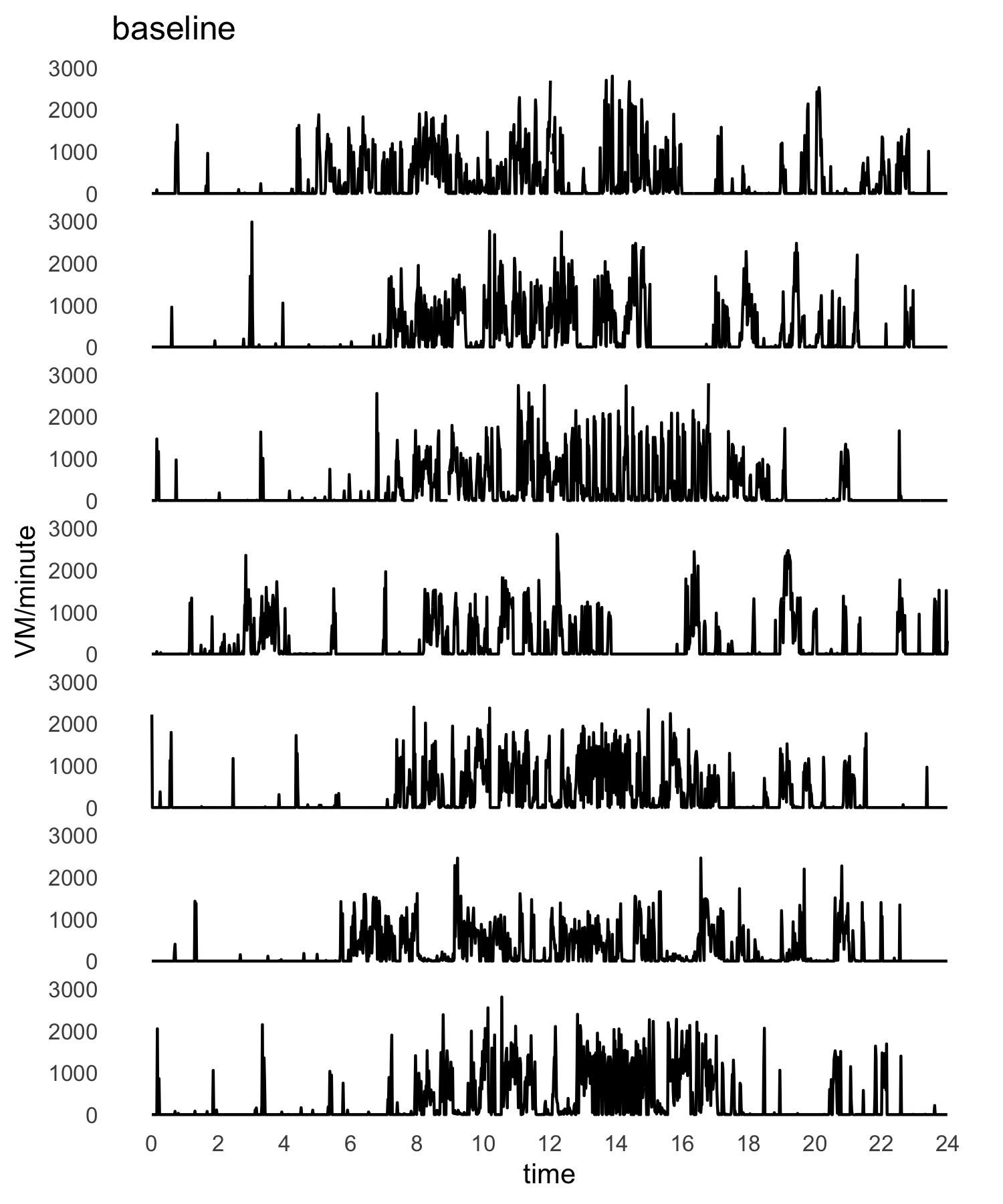}
    \captionsetup{justification=centering} 
    \caption{Activity count during multiple days at baseline}
    \label{subfig:V0}
  \end{subfigure}
   \begin{subfigure}[b]{0.5\textwidth}
    \includegraphics[width=\linewidth,height=0.41\textheight, keepaspectratio]{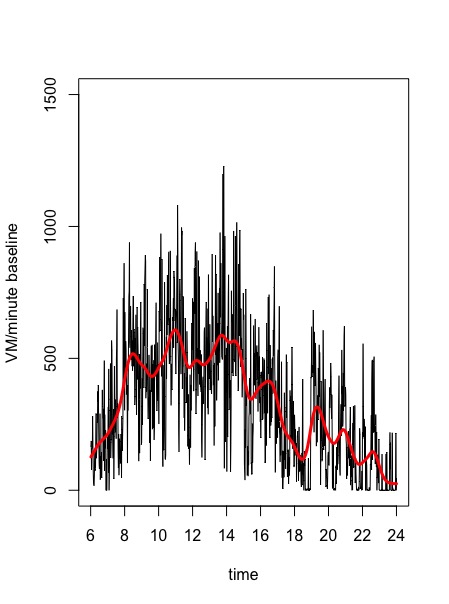}
    \captionsetup{justification=centering} 
    \caption{Mean and smoothed mean curve (red) at baseline (6 a.m. to midnight)}
    \label{subfig:V0_smoothed}
  \end{subfigure}
  
   \begin{subfigure}[b]{0.5\textwidth}
    \includegraphics[width=\linewidth,height=0.41\textheight, keepaspectratio]{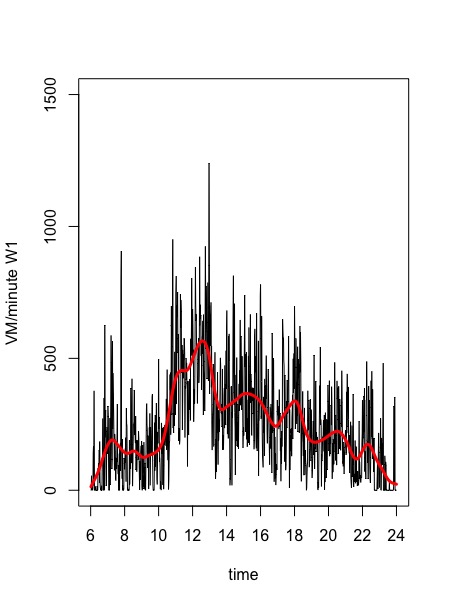}
    \captionsetup{justification=centering} 
      \caption{Mean and smoothed mean curve (red) at W1 (6 a.m. to midnight)}
      \label{subfig:V1_smoothed}
  \end{subfigure}  
   \begin{subfigure}[b]{0.5\textwidth}
    \includegraphics[width=\linewidth,height=0.41\textheight, keepaspectratio]{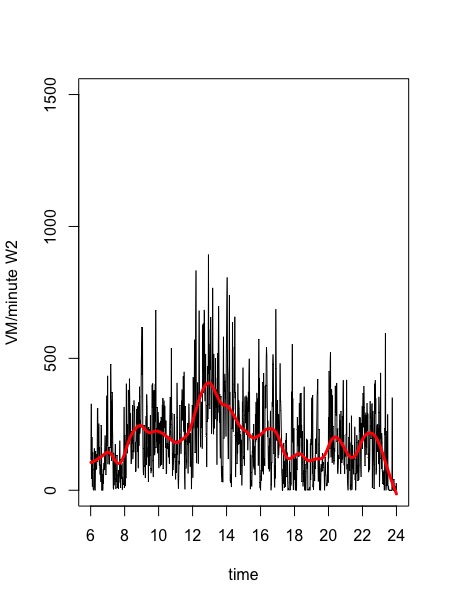}
    \captionsetup{justification=centering} 
       \caption{Mean and smoothed mean curve (red) at W2 (6 a.m. to midnight)}
       \label{subfig:V2_smoothed}
  \end{subfigure}
  
  \caption{Example of one participant's activity count and smoothed means}
  \label{fig:VM_smooth}
\end{figure}

The retained VM for each participant were averaged across days within each visit, yielding a single diurnal activity profile per participant per visit. VM counts were aligned to chronological time ($t = 1, \dots, 1080$, representing minutes from 6 a.m. to midnight), and a cubic smoothing spline (25 degrees of freedom) was applied at each participant-visit to reduce measurement noise. Figure~\ref{fig:VM_smooth} illustrates one participant's PA profile from daily minute-level measurements (\ref{subfig:V0}) to the smoothed curve (\ref{subfig:V0_smoothed}) at baseline, with smoothed curves at W1 (\ref{subfig:V1_smoothed}) and W2 (\ref{subfig:V2_smoothed}).

The grand mean and standard deviation (SD) of the smoothed PA counts were computed pooling across all participants and visits. Each participant--visit curve was then centered by this grand mean and scaled by $4 \times \text{SD}$ so that the majority of scaled curves lie approximately within $[-1, 1]$. The time domain $[1, 1080]$ was likewise rescaled to $[-1, 1]$, balancing the penalty weights on deformation energies along the amplitude and time axes \citep{zou2023riemann}. For each participant at each visit, the signed Area Under the scaled PA Curve (AUC) was defined as
$\text{net-AUC} \;=\; \int_{\{t:\, y(t) > 0\}} y(t)\, dt \;-\; \int_{\{t:\, y(t) < 0\}} \lvert y(t) \rvert\, dt$,
where $y(t)$ denotes the scaled PA curve. Although this reduces to the ordinary integral $\int y(t)\, dt$ that can be estimated by trapezoid rule, the decomposition emphasizes the directional interpretation. The between-visit difference, $\Delta\text{net-AUC}^{(\eta)}$, then served as a signed summary of net PA change, where $\eta \in \{0, 1\}$ indexes the transition period ($\eta = 0$: baseline to W1; $\eta = 1$: W1 to W2). Positive values indicate an increase in overall PA; negative values indicate a decrease.

\subsection{Physical function outcome}
Physical function (PF) reflects the capacity to perform daily activities requiring locomotion, balance, and muscular endurance, and its decline is a well-established precursor to disability, loss of independence, and mortality in older adults \citep{guralnik1995lower,studenski2011gait}.
Importantly, low PF is a risk factor for CVD \citep{stefanick2016relationship,yazdanyar2014association} and predicts incident CVD events after adjustment for traditional risk factors \citep{saquib2019changes,bellettiere2020short}. There is also compelling evidence that PA is central to preserving PF and mobility in later life \citep{life2006effects,pahor2014effect,PAG_Advisory_Committee_Report_2018,king2001interventions}. Given its dual role as a key pathway through which PA influences aging-related outcomes and as an independent predictor of incident CVD, PF served as the primary outcome of this study.

PF was assessed using RAND-36 Physical Functioning subscale, which quantifies health-related limitations in daily physical activities on a 0--100 scale, with higher values indicating better functioning \citep{hays1993rand}. Because participants' RAND-36 assessments and accelerometer wear wave did not always occur simultaneously, the first day of accelerometer wear at each of W1 and W2 was matched to the nearest RAND-36 assessment within a $\pm$190-day window. This window was chosen to minimize the time discrepancy between two different instruments while maximizing participant retention.
\subsection{Covariates}
Baseline covariates were selected based on prior literatures and clinical relevance \citep{nguyen2024prospective,LaCroix:2019eb}. With the exception of race/ethnicity and education, which were obtained at the WHI baseline, all others were collected at the OPACH baseline. Questionnaires assessed age, race/ethnicity (Black, White, Hispanic/Latina), education (high school equivalent or less, some college, college graduate or higher), alcohol consumption (nondrinker, $<$ 1 drink/week,  $\ge$1 drink/week, unknown), current smoking status (no, yes), and self-rated general health (excellent/very good, good, poor/very poor).
Moderate-to-vigorous PA (MVPA) was defined using the OPACH Calibration Study VM count cut-point ($\ge$519 counts per 15 seconds) \citep{evenson2015calibrating}. Baseline average daily MVPA (minutes/day) was computed over awake wear time by summing MVPA across valid days and averaging over total number of valid days.

Multimorbidity was categorized as 0, 1, 2, or $\ge$ 3 of 12 prevalent conditions: history of CVD, cancer, diabetes, hip fracture, osteoarthritis, depression, chronic obstructive pulmonary disease, cognitive impairment, self-reported moderate/severe vision or hearing loss, $\ge$2 falls in past 12 months, and urinary incontinence \citep{rillamas2016impact}. Study staff measured systolic and diastolic blood pressure (SBP and DBP; mmHg) by auscultation using an aneroid sphygmomanometer, with participants seated following 5 minutes of quiet rest; 2 measurements were averaged. Height and weight were assessed by tape measure and bathroom scale to calculate BMI (kg/m$^2$). Fasting (12-hour) blood samples were collected close in time ($\le$6 months) to the accelerometer wear interval. Serum glucose, insulin, total cholesterol, high-density lipo-protein cholesterol (HDL-C), and high-sensitivity C-reactive protein (hsCRP) were measured using standardized Clinical Laboratory Improvement Act–approved methods at the University of Minnesota Fairview Advanced Research and Diagnostics Laboratory \citep{lamonte2017both}. The Reynolds Risk Score, a summary CVD risk score representing the predicted 10-Year probability of having a clinical CVD event, was calculated using age, SBP, hsCRP, total and HDL cholesterol, diabetes, smoking, and family history of myocardial infarction \citep{ridker2007development}.

\section{Statistical analyses} \label{sec:analysis}
\subsection{Deformation-based representation of change in PA}\label{RM_momenta}
We modeled longitudinal \emph{change} in diurnal PA patterns using the Riemannian diffeomorphic framework of \citet{zou2023riemann}. 
Participant's daily PA pattern was represented as a smooth curve at each visit, the longitudinal change from two consecutive visits (e.g, baseline -- W1, W1 -- W2) were modeled as a Riemann Manifold diffeomorphism, a smooth transformation that deforms the source curve into the target curve. 

Formally, let $C_i^{(q)}$ denote the $i^{th}$ participant's smoothed PA curve at
visit $q$. The change from visit $q{-}1$ to visit $q$ is modeled as
\begin{equation}\label{eq:deformation}
  C_i^{(q)} = \gamma\!\bigl(\boldsymbol{v}_i^{(q)},\;\cdot\;\bigr)
               \circ\, C_i^{(q-1)},
\end{equation}
where $\gamma(\boldsymbol{v}_i^{(q)}, \cdot)$ is a subject-specific
diffeomorphism that maps the source curve $C_i^{(q-1)}$ to the target curve
$C_i^{(q)}$, and $\circ$ denotes function composition.

The diffeomorphism is governed by a time-varying vector field
$\boldsymbol{v}_i^{(q)}$ embedded in a reproducing kernel Hilbert space (RKHS) equipped with a Gaussian isotropic kernel
\citep{charlier2017fshape,zou2023riemann}. The deformation evolves over an
auxiliary parameter $\tau \in [0,1]$, where $\tau = 0$ corresponds to the source
curve and $\tau = 1$ to the target. Intuitively, the vector field
$\boldsymbol{v}_\tau$ can be viewed as a series of forces that continuously
deform the PA curve from one visit to the next at each infinitesimal step
$\tau$. Crucially, these forces operate simultaneously in both the amplitude
($y$) and temporal ($x$) domains, enabling the model to capture changes in PA
magnitude (e.g., an overall increase or decrease in activity) and temporal/phase
shifts (e.g., redistribution of PA bouts to earlier or later hours) within
a single unified framework.

\begin{figure}[h]
     \includegraphics[width=\textwidth]{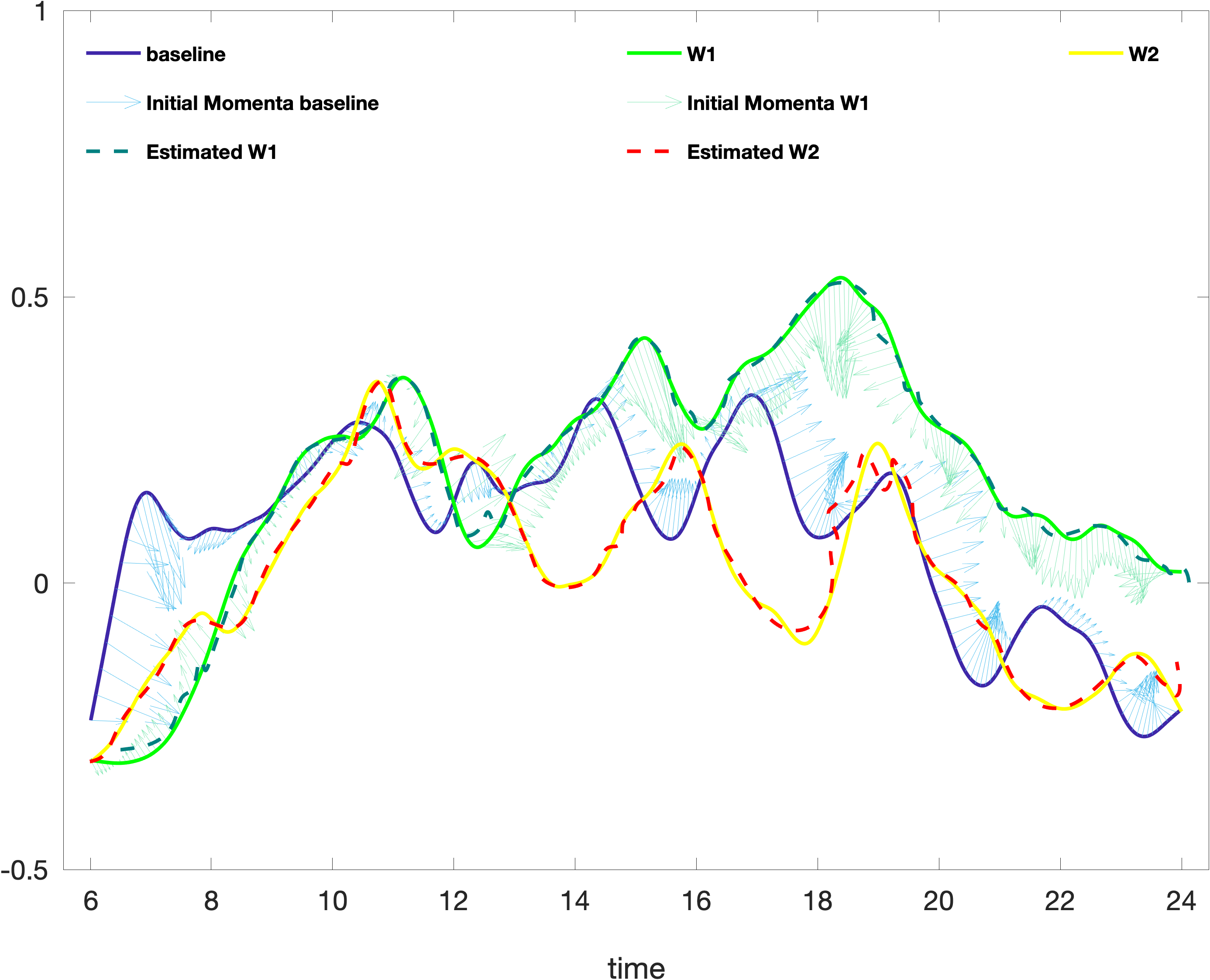}
     \caption{One participant's PA curves deformation and initial momenta among visits. The solid curves in blue, green and yellow show this participant's observed (smoothed and scaled) PA curves at baseline, W1 and W2. Light blue arrows are the initial momenta from baseline to W1 and light green arrows are the initial momenta from W1 to W2. Two dashed curves in dark green and red are the estimated PA curves at W1 and W2, respectively. The arrows of momenta depict the changes in PA magnitude (up/down) and temporal shift (left/right) simultaneously. Close agreement between the dashed (estimated) and solid (observed) curves indicates good reconstruction of the observed visit-to-visit change.}
     \label{fig:subj_deformaton}
  \end{figure}

Subject to the constraint on \emph{deformation energy},
$\int_0^1 \|\boldsymbol{v}_\tau\|^2\, d\tau$, the deformation in Eq 
\eqref{eq:deformation} is unique and fully determined by the \emph{initial
momenta}: the values of the vector field at $\tau = 0$ evaluated at a set of
preselected control points on the curve. In our analysis, the control
points are selected as the $1{,}080$ minutes from
6:00~a.m.\ to midnight. At each minute $p$, the initial momentum is a
two-dimensional vector,
\begin{equation*}
  \boldsymbol{m}_{i,p}^{(\eta)}
    = \bigl(m_{i,p,x}^{(\eta)},\; m_{i,p,y}^{(\eta)}\bigr),
\end{equation*}
where the $x$-component encodes the local temporal shift and the $y$-component
encodes the local amplitude change; $\eta \in \{0, 1\}$ corresponds to  transition periods (baseline -- W1 and W1 -- W2). 
The optimal initial momenta are calculated by minimizing a penalized objective that balances the similarity
of the deformed source curve to the observed target against total deformation
energy, solved via Hamiltonian optimization using the \texttt{fshapesTK} package \citep{charlier2017fshape}.
For participant \(i\) and period \(\eta\), we summarize the overall amount of deformation by the initial deformation energy,
$d_i^{(\eta)} = \sum_{p=1}^{1080} \left[\bigl(m^{(\eta)}_{i,p,x}\bigr)^2 + \bigl(m^{(\eta)}_{i,p,y}\bigr)^2\right]$.
Details of initial momentum estimation, deformation energy and optimization can be found in previous publications \citep{charlier2017fshape,zou2023riemann}. As an example, Figure~\ref{fig:subj_deformaton} showed the estimated deformations and initial momenta of one participant from the OPACH-WHISH sub-cohort.

\subsection{Multivariate functional principal component analysis on changes in PA} \label{stats_mfpca}
To extract the dominant modes of longitudinal PA change and to preserve the
distinction between these two clinically interpretable $x$ (Temporal) and $y$ (Amplitude) domains, we applied
multivariate functional principal component analysis (MFPCA) using the approach
of \citet{happ2018multivariate}. 

In this framework, univariate eigenfunctions and scores were first estimated separately on each domain, and the multivariate eigenfunctions and scores were obtained as linear combinations of their univariate counterparts with a weight factor. The estimation steps were described in more details in Supplemental (\ref{MFPCA_steps}) and \citep{happ2018multivariate}. 
MFPCA was applied separately to each transition period with identical parameter settings. 

An alternative, simpler strategy is to concatenate the $x$- and $y$-components of the initial momenta into a single vector function and apply standard univariate FPCA (UFPCA) \citep{fdapace}. 
If the measurement on both domains had identical grids (time-points), comparable variance, and the principal modes were not highly correlated across domains, concatenated UFPCA can approximate MFPCA reasonably. Otherwise, naive concatenation
risks producing blended components in which temporal-shift modes are absorbed into amplitude modes (or vice versa), obscuring domain-specific interpretability. We therefore report results from both approaches and quantify their differences.

\subsection{Association between changes in PA and health outcomes}\label{method::association}
Three sets of predictors were derived in earlier sections to characterize
longitudinal changes in diurnal PA patterns: (i)~MFPCA principal component
scores (Section~\ref{stats_mfpca}), which capture the dominant modes of change
in both temporal shift and amplitude; (ii)~$\Delta\text{net-AUC}^{(\eta)}$
(Section~\ref{qc_process}), which summarizes net directional change (positive versus negative) in PA without information on temporal shift, and (iii)~initial deformation energy $d_i^{(\eta)}$ (Section~\ref{RM_momenta}), which summarizes overall amount of PA change without directionality in timing and amplitude.

Because PC1 and $\Delta$net-AUC were highly collinear in both periods (Pearson $r=0.9$), they were not included in the same model. In contrast, deformation energy showed weak correlations with PC1 ($|r|\le 0.15$) and with $\Delta$net-AUC ($|r|\le 0.30$) in both periods, so deformation energy was modeled jointly with PC1 and, separately, with $\Delta$net-AUC. Two LMMs were carried out with participant-specific random effects to account for repeated PF outcome at W1 and W2: \textbf{Model 1} included the stacked deformation-based change features (PC scores) and deformation energy, together with baseline PF scores and other covariates; \textbf{Model 2} replaced the PC scores with the scalar change summary, $\Delta$net-AUC, while retaining deformation energy, baseline PF and covariates. Inclusion of interaction terms between period and the predictors were determined using likelihood ratio tests (LRTs) comparing nested models. To reduce overfitting and improve interpretability, we applied LASSO \citep{groll2014variable} to select covariates and PCs in the LMM via R package \texttt{glmmLasso}\citep{glmmLasso}. The LASSO penalty parameter was determined by Bayesian Information Criterion (BIC), and the significance threshold for two LMMs was adjusted to $\alpha=0.025$ by Bonferroni method.

All statistical analyses were conducted in the R statistical programming environment (version 4.3.1) \cite{rcoreteam2023}, except for the initial momenta estimation and optimization steps, which were implemented using the MATLAB package \texttt{fshapesTK} \cite{charlier2017fshape} and executed on the National Research Platform Nautilus Kubernetes Cluster \citep{Nautilus}. R package \texttt{nlme} for LMM is available from~\citep{Pinheiro2023} and package \texttt{MFPCA} is available from \citep{MFPCA:Clara}.

\section{Results}\label{sec:results} 
Table~\ref{tab:baseline} summarizes baseline characteristics of the OPACH--WHISH sub-cohort. Continuous variables are reported as means (SD), and categorical variables as counts (n) and percentages (\%). At baseline, the mean age was 77 years and the mean body mass index (BMI) was 27.8~kg/m$^2$. More than half of participants were White, 45\% has attained at least a college degree, and approximately 60\% rated their general health as excellent. The mean MVPA was 61~minutes/day, and the mean PF score was 75.

\begin{table}[ht]
\centering
\caption{Baseline Characteristics of OPACH-WHISH sub-cohort ($N^+$ = 1,730)}\label{tab:baseline}
\begin{tabular}{lclc}
\hline
\textbf{Continuous} & \textbf{Mean (SD)} &
\textbf{Categorical} & \textbf{n$^1$ (\%)} \\
\hline
Age			& 77.2 (6.55)	& Race/ethnicity	&  \\
(year)		&			& \quad White		& 897 (51.8)\\
			&			& \quad Black		& 560 (32.4)\\		
			&			& \quad Hispanic/Latina & 273 (15.8)  \\\\

BMI			& 27.8 (5.45)	& Education				&  \\
(kg/m$^2$)	&			& \quad High school or less	& 294 (17.1)\\
			&			& \quad Some college		& 658 (38.2)\\		
			&			& \quad College graduate or higher & 769 (44.7)  \\\\
				
MVPA		& 60.6 (37.05)	& Alcohol consumption			&  \\
(minutes/day)	&			& \quad Nondrinker 				& 572 (33.1) \\
			&			& \quad $<$ 1 drink/week 	& 570 (32.9) \\	
			&			& \quad $\ge$1 drink/week		& 521 (30.1) \\		
			&			& \quad Unknown 				& 67 (3.9)  \\\\
				
RAND-36  	& 75.1 (23.02)	& No. of chronic conditions&  \\
PF Score		&			& \quad None			& 386 (22.4) \\
			&			& \quad 1				& 660 (38.2)\\		
			&			& \quad $\ge$2 		& 681 (39.4)\\\\
				
Reynolds 		& 10.1 (8.79)	& Current smoke status	&  \\
Risk Score	&			& \quad No			& 1628 (97.9) \\
			&			& \quad Yes 			& 35 (2.1)\\\\
					
Log (hsCRP)  	& 0.6 (1.01)	& Self-rated health	&  \\
(mg/L)		&		& \quad Excellent 		& 1049 (60.7)\\
			&			& \quad Good		&  605 (35.0)\\		
			&			& \quad Poor & 74 (4.3) \\\\

Insulin (pmol/L) & 84.5 (98.55)	& 	&  \\
Glucose (mg/dL) & 96.0 (22.97)	& 	&  \\
Blood Pressure (mmHg)  & 	& 	&  \\
\quad SBP         & 124.5 (13.97) 	& 	&  \\
\quad  DBP        & 72.1 (8.27)	& 	&  \\
Cholesterol (mg/dL) & 	& 	&  \\
\quad Total         & 199.3 (38.01) 	& 	&  \\
\quad  HDL-C     &  61.8 (15.53)	& 	&  \\\\
\hline
\multicolumn{3}{l}{\footnotesize $^1$ n of categories may not sum up to 1,730 due to missing values of baseline covariates} \\
\end{tabular}
\end{table}

\subsection{Deformation-based principal components of diurnal PA change}
At both periods, baseline--W1 and W1–W2, 15 PCs are required in MFPCA to explain at least 90\% of the variability, with PC1 accounting for the largest share among these components (22.4\% and 20.8\%, respectively), followed by PC2 (10.9\% and 11.5\%). Detailed variance explained by each PC in both periods is reported in Supplemental (\ref{varExp}). 

Figure \ref{fig:MFPCA_mu} presents the MFPCA-estimated mean functions (top row) and corresponding deformations (bottom row) for the $x$ (temporal shift) and $y$ (magnitude shift) domains. In both periods, the mean $x$-domain function cross zero near 10:20 a.m., transitioning from rightward (positive) to leftward (negative) temporal shifts (\ref{subfig:meanX}). The mean $y$-domain function remains negative throughout, indicating persistent downward vertical shifts, with the deepest drop occurring at 7:52 p.m. in baseline–W1 (\ref{subfig:meanY}).

The bottom panels illustrate how the estimated initial momenta deform one visit's smoothed PA curves toward the follow-up visit's curves. In the baseline–W1 panel (\ref{subfig:mean_deformV0V1}), arrows point rightward and downward before 10:20 a.m. and leftward and downward thereafter, dragging the observed baseline mean (blue) toward W1 (green); the momentum-derived estimate mean PA curve (dashed red) closely follows the observed W1 mean. The W1–W2 panel (\ref{subfig:mean_deformV1V2}) exhibits a similar deformation pattern but with attenuated temporal and vertical shifts, consistent with the shorter average inter-visit interval (one year for W1–W2 versus three years for baseline–W1), and may also reflect smaller change in PA patterns at older ages.

Figure~\ref{fig:MFPCA_efunc1} presents the PC1 eigenfunction (top row) and its corresponding deformation (bottom row). 
Temporal shifts of PA under PC1 are similar across both periods (\ref{subfig:pc1X}): predominantly leftward before 10:00 a.m. (negative eigenfunction $x$-coordinates) and rightward afterward (positive $x$-coordinates). Magnitude changes in PA are also consistent across periods and primarily upward (positive $y$-coordinates) (\ref{subfig:pc1Y}). The $y$-coordinate change is more pronounced than the $x$-coordinate change, indicating that PC1 captures more variability in PA magnitude than in timing. As with any PC, this mode is bidirectional: the plotted eigenfunction represents one direction of PC1, so a positive PC1 score corresponds to change in the displayed direction and a negative score to the opposite. These changes are visualized more directly in the bottom deformation panels (\ref{subfig:pc1_V0V1}, \ref{subfig:pc1_V1V2}): blue curves (identical to those in Figure~\ref{subfig:mean_deformV0V1},~\ref{subfig:mean_deformV1V2}) represent the observed sample mean curves at the earlier visit of each period (baseline and W1, respectively), and dashed red curves show the estimated curves deformed from the blue curves following PC1's initial momenta. 
Because PC1 explains the largest portion of variability after removal of the mean function, it captures the residual primary mode of diurnal PA change, with positive scores indicating increase and negative indicating decrease in PA, especially between 10 a.m. and 7 p.m. 

Deformation panels of PC2 are shown in Figure (\ref{fig:MFPCA_efunc2_deform}). Both temporal shifts and magnitude changes are consistent across the two periods. PC2 characterizes a leftward shift and localized increase in morning activity for positive scores (or a decrease for negative scores), illustrated by the up-and-leftward-pointing arrows from 6 a.m. to around 11 a.m.\ (\ref{subfig:pc2_V0V1_deform},~\ref{subfig:pc2_V1V2_deform}), followed by a slight downward shift thereafter.

Additional eigenfunctions and deformation figures are provided in Supplemental (\ref{otherPCs}), revealing distinct patterns of PA change during morning, afternoon, and evening phases. Beyond the top four PCs, each additional component explains only a small proportion of variability and PA changes associated with those PCs are minimal. 

Importantly, in this OPACH–WHISH subcohort, the eigenfunctions and associated deformations of PA patterns for the top four PCs show similar patterns across the two periods. This cross-period similarity makes it reasonable to treat, for example, PC1 from baseline–W1 and PC1 from W1–W2 as representing the same underlying mode of PA change. Accordingly, the corresponding PC scores can be stacked across the two periods and included as a common predictor in the subsequent LMM association analyses.

\begin{figure}[h]
  \centering
  
 \begin{subfigure}[b]{0.48\textwidth}
    \centering
    \includegraphics[width=\textwidth]{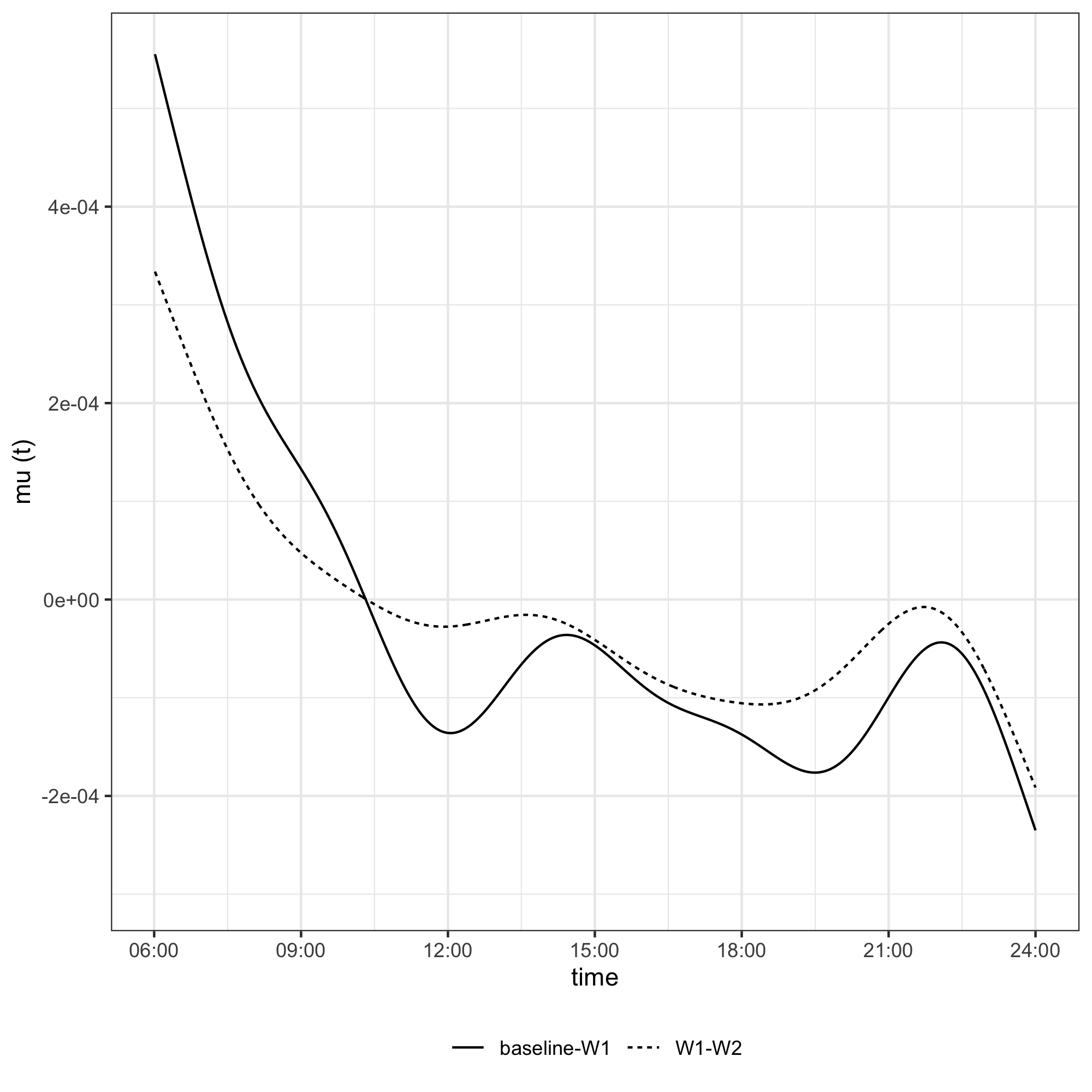}
    \captionsetup{justification=centering} 
    \caption{Mean of initial momenta in the $x$ domain}
    \label{subfig:meanX}
  \end{subfigure}
  \hspace{0.02\textwidth} 
  \begin{subfigure}[b]{0.48\textwidth}
    \centering
    \includegraphics[width=\textwidth]{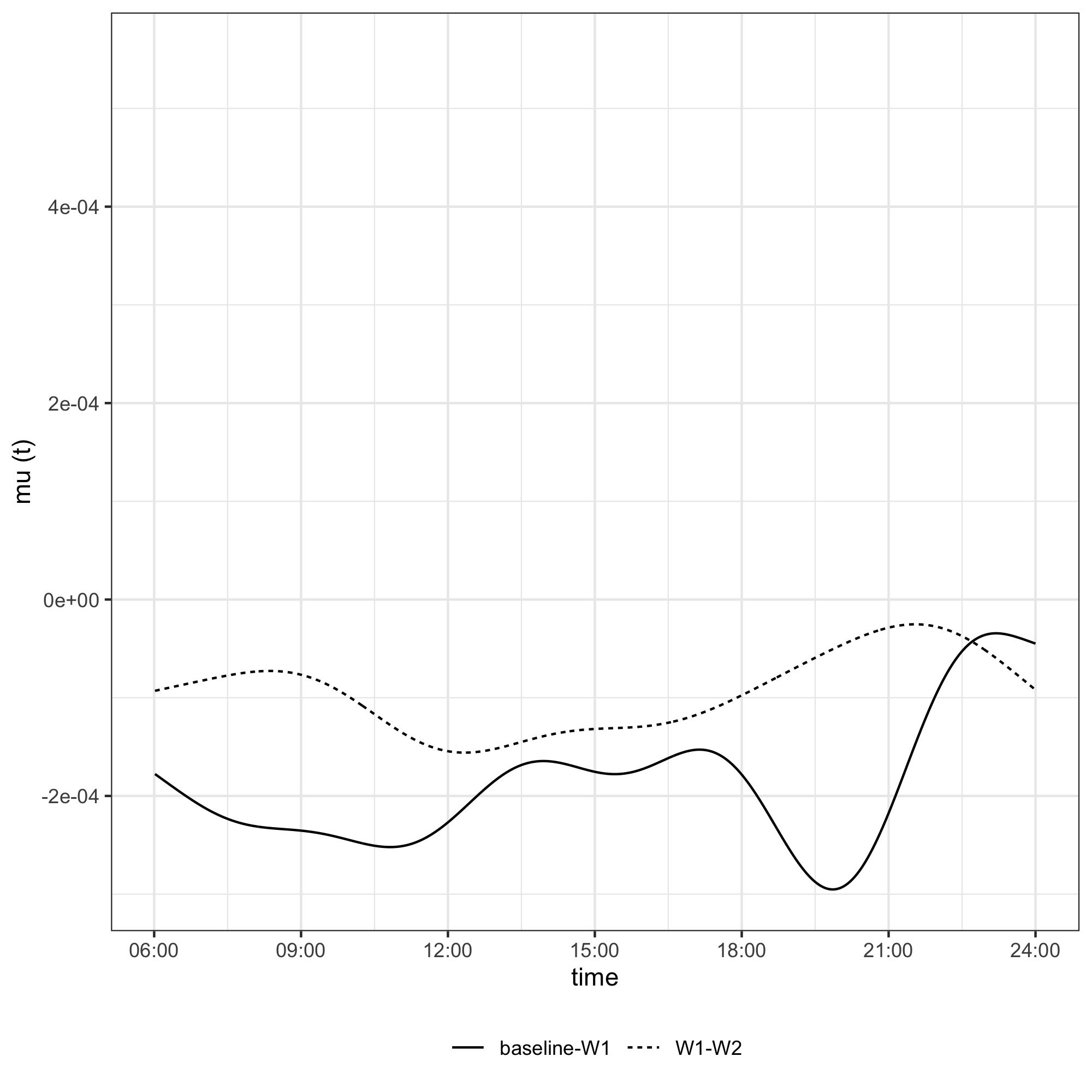}
    \captionsetup{justification=centering} 
      \caption{Mean of initial momenta in the $y$ domain}
      \label{subfig:meanY}
  \end{subfigure}

  \vspace{1em} 
  
  \begin{subfigure}[b]{0.48\textwidth}
    \centering
    \includegraphics[width=\textwidth]{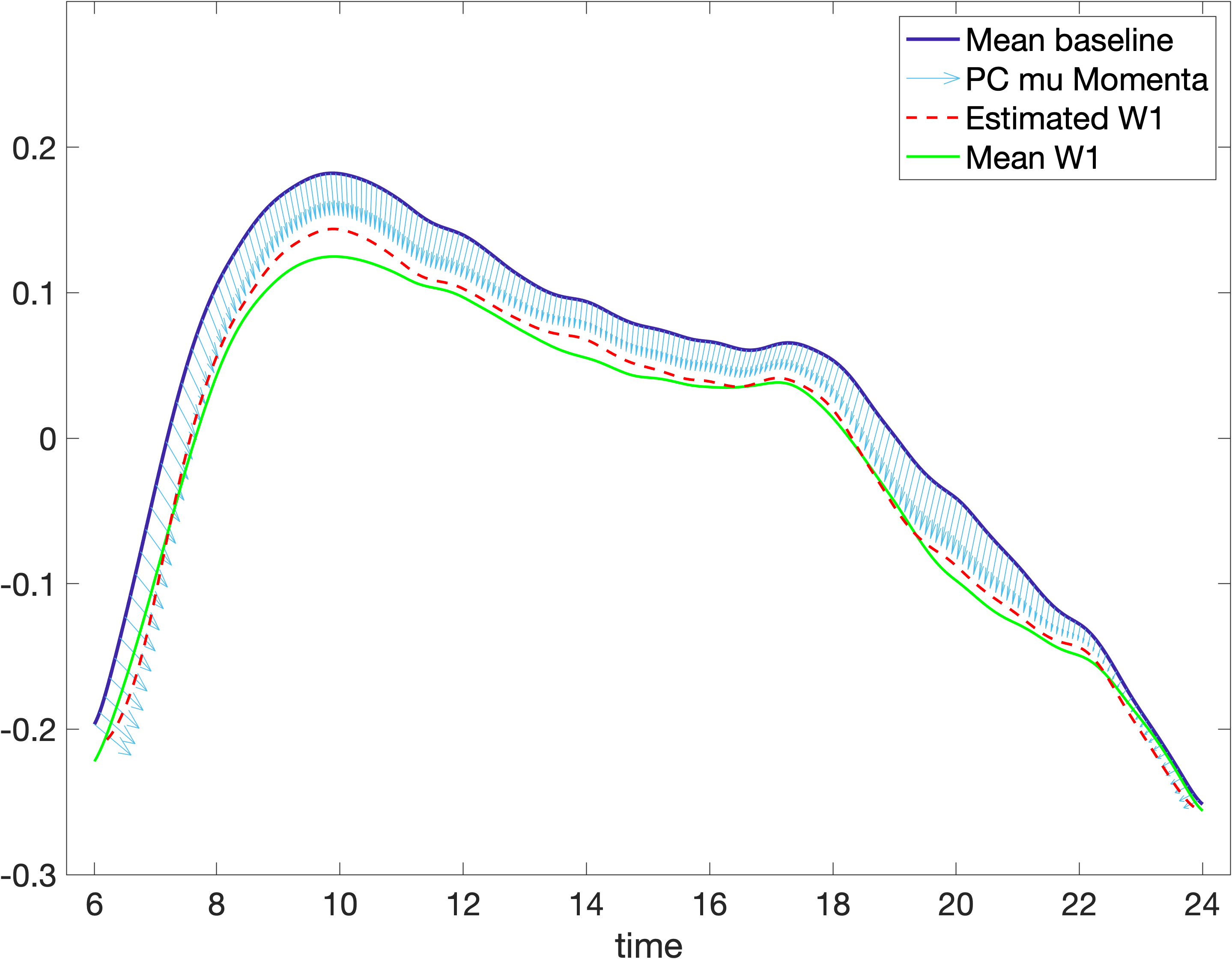}
    \captionsetup{justification=centering} 
    \caption{Mean of initial momenta and corresponding deformation from baseline to W1.}
    \label{subfig:mean_deformV0V1}
  \end{subfigure}
  \hspace{0.02\textwidth} 
  \begin{subfigure}[b]{0.48\textwidth}
    \centering
    \includegraphics[width=\textwidth]{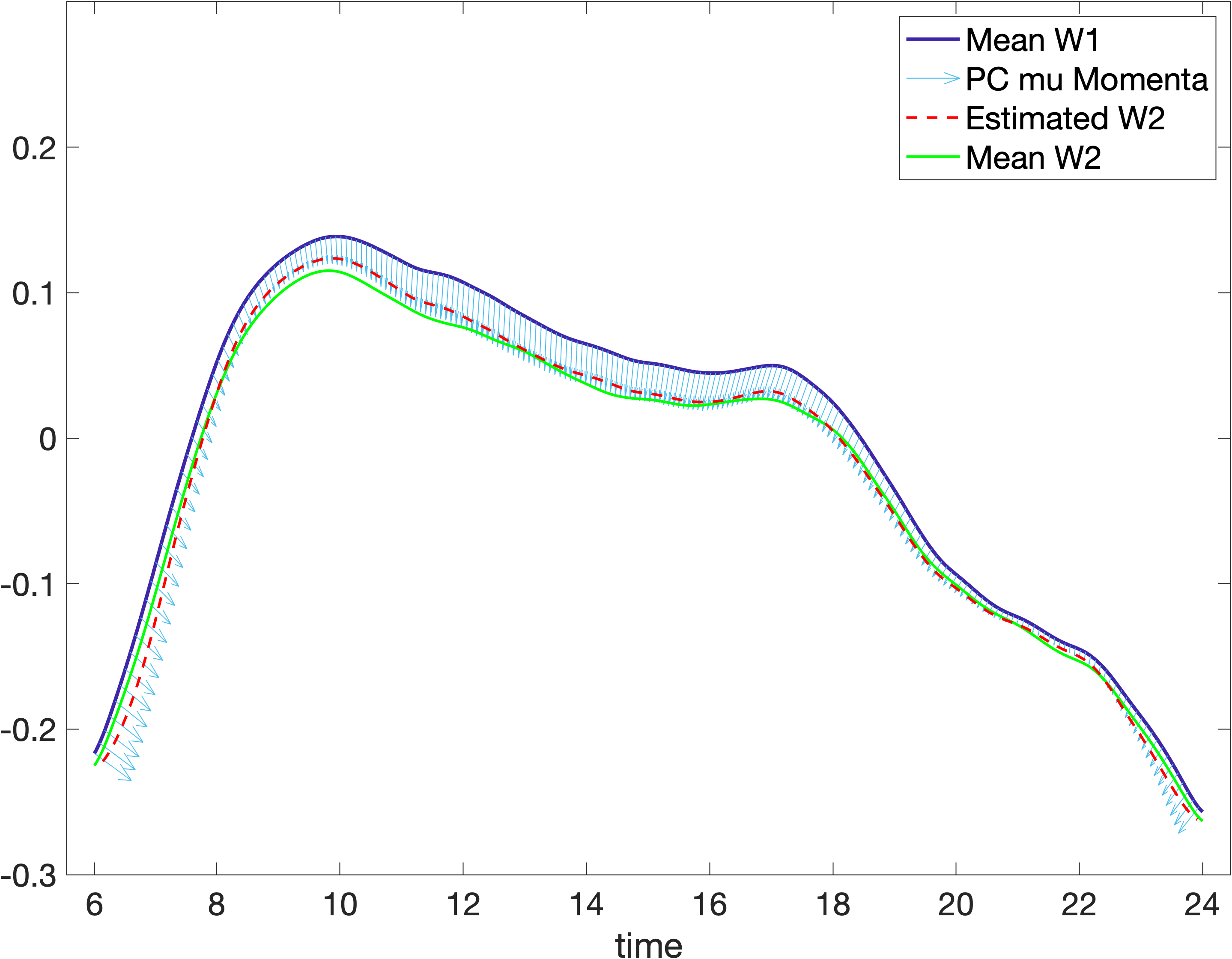}
    \captionsetup{justification=centering} 
    \caption{Mean of initial momenta and corresponding deformation from W1 to W2  \vspace{2.8ex}}
    \label{subfig:mean_deformV1V2}
  \end{subfigure}

  \caption{MFPCA mean functions and deformations}
  \label{fig:MFPCA_mu}
\end{figure}

\begin{figure}[h]
  \centering

  \begin{subfigure}[b]{0.48\textwidth}
    \centering
    \includegraphics[width=\textwidth]{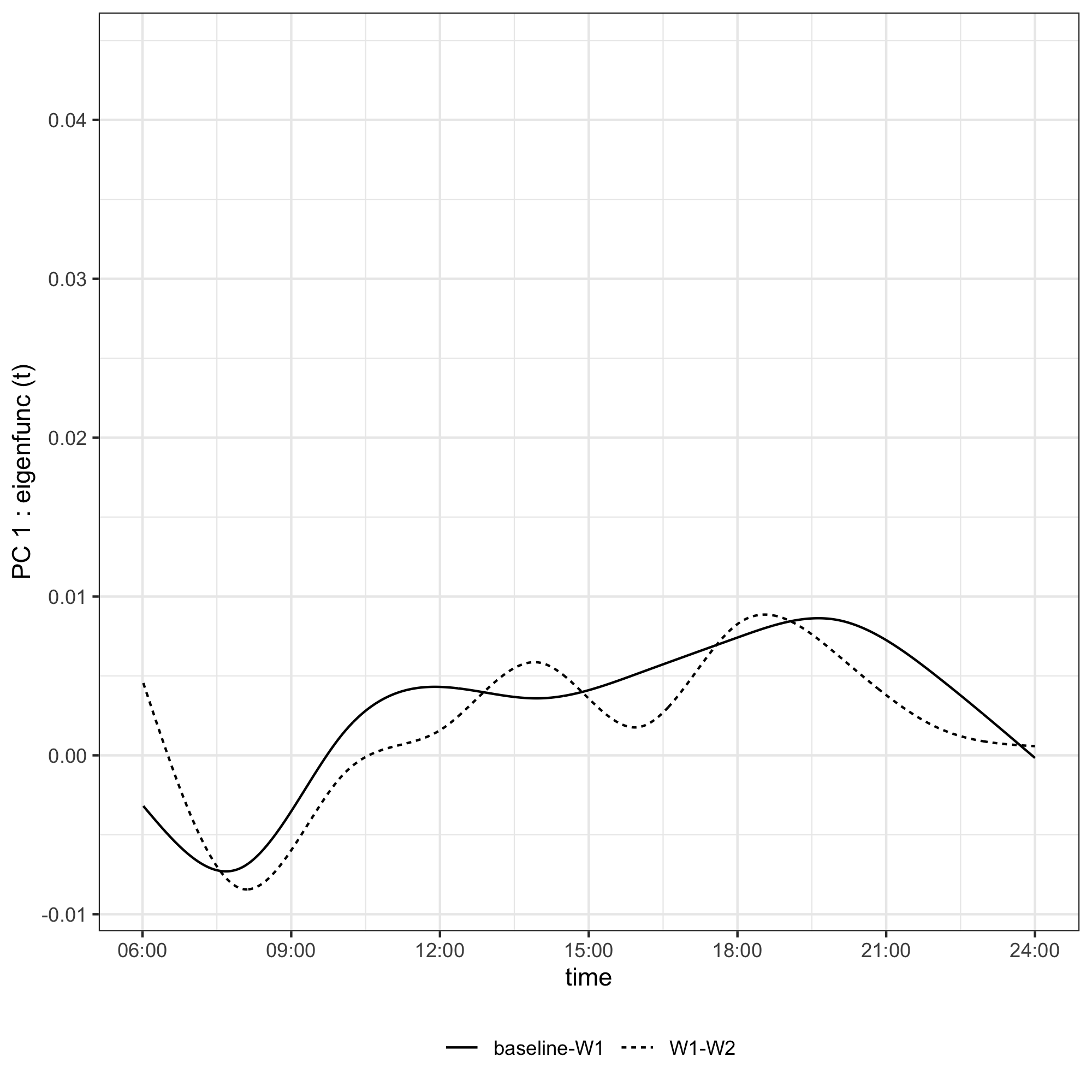}
     \captionsetup{justification=centering} 
    \caption{PC1 eigenfunction in $x$ domain }
    \label{subfig:pc1X}
  \end{subfigure}
  \begin{subfigure}[b]{0.48\textwidth}
    \centering
    \includegraphics[width=\textwidth]{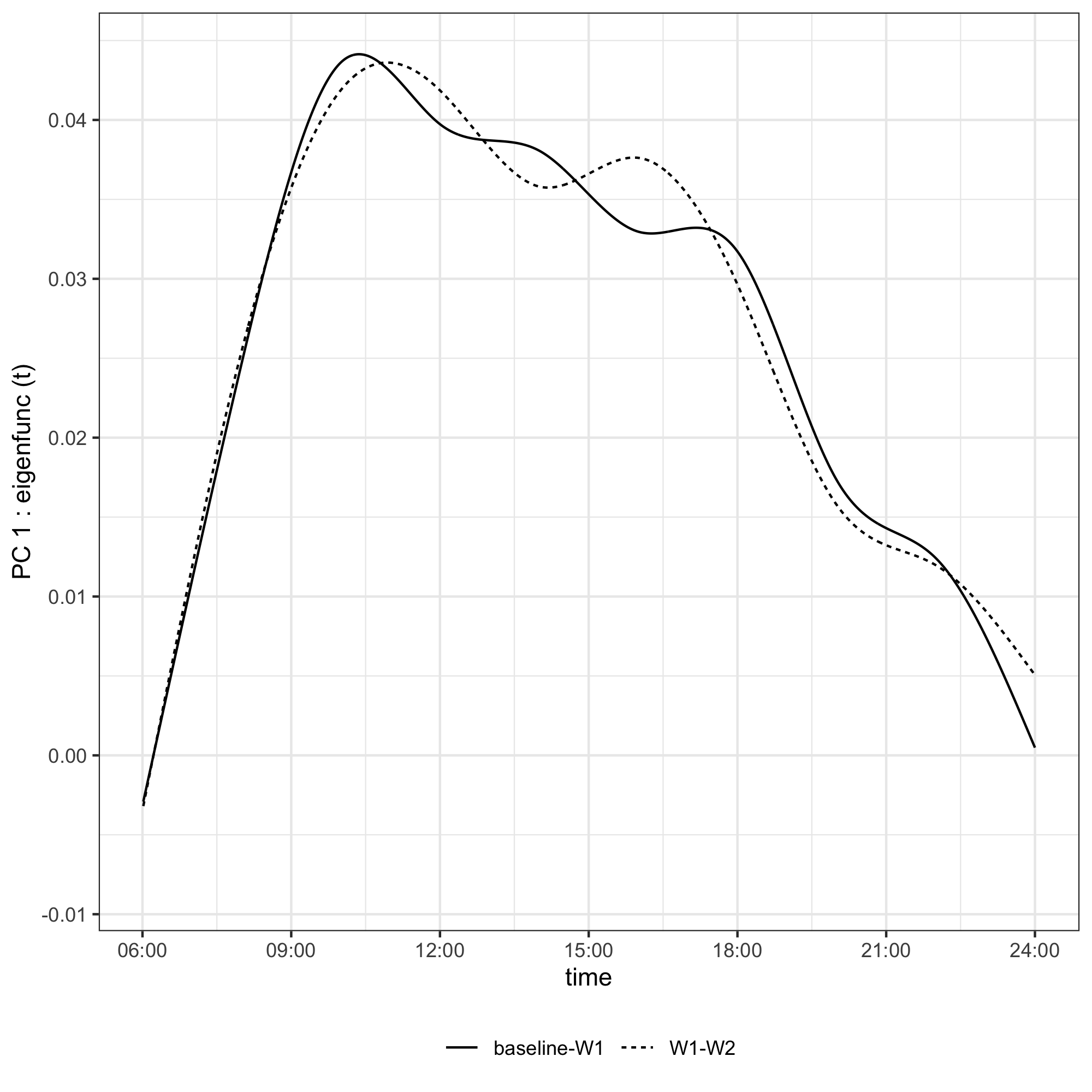}
    \captionsetup{justification=centering} 
    \caption{PC1 eigenfunction in $y$ domain}
    \label{subfig:pc1Y}
  \end{subfigure}

  \vspace{1em}

    \begin{subfigure}[b]{0.48\textwidth}
    \centering
    \includegraphics[width=\textwidth]{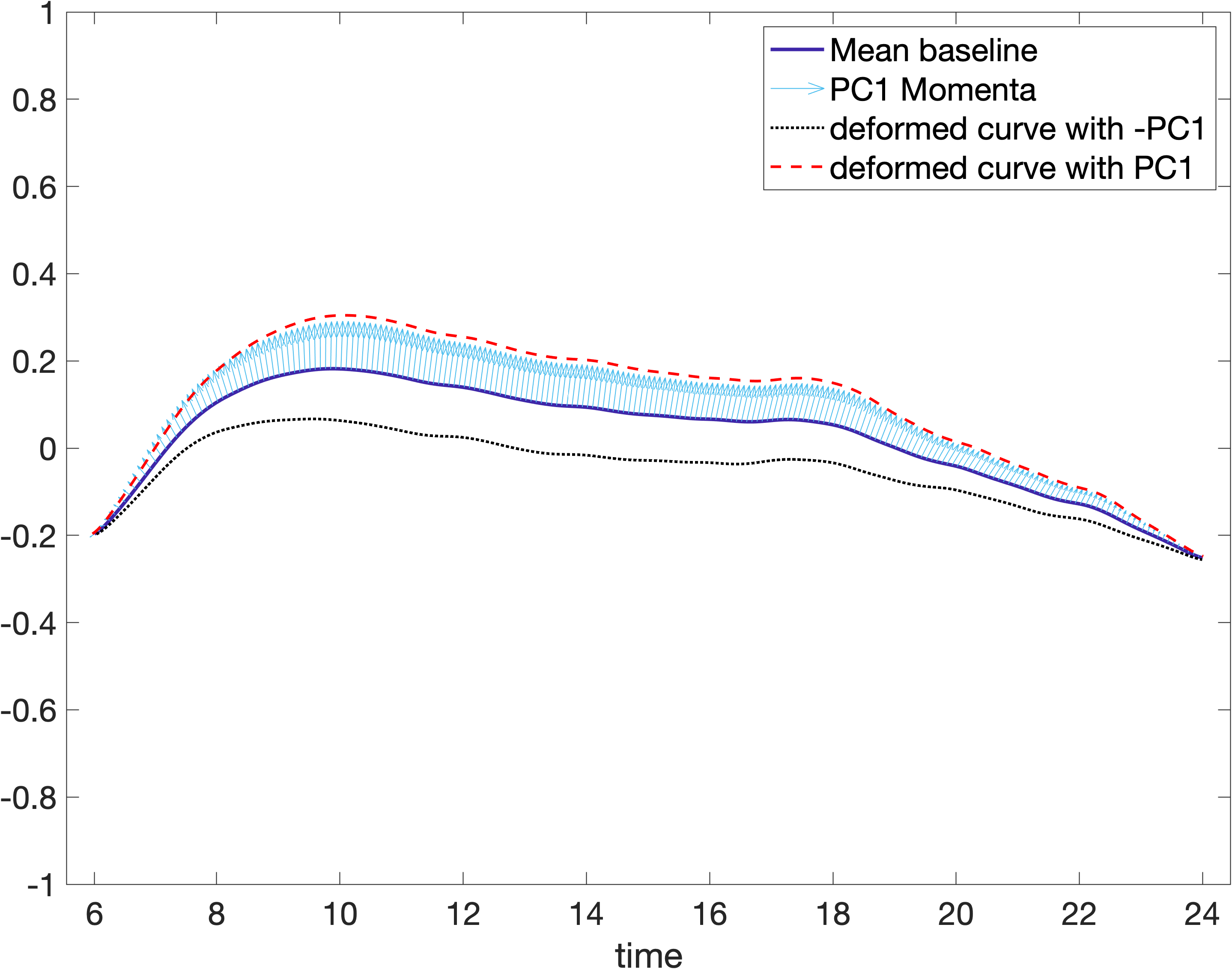}
    \captionsetup{justification=centering} 
    \caption{PC1 initial momenta and deformations in period baseline--W1}
    \label{subfig:pc1_V0V1}
  \end{subfigure}
  \begin{subfigure}[b]{0.48\textwidth}
    \centering
    \includegraphics[width=\textwidth]{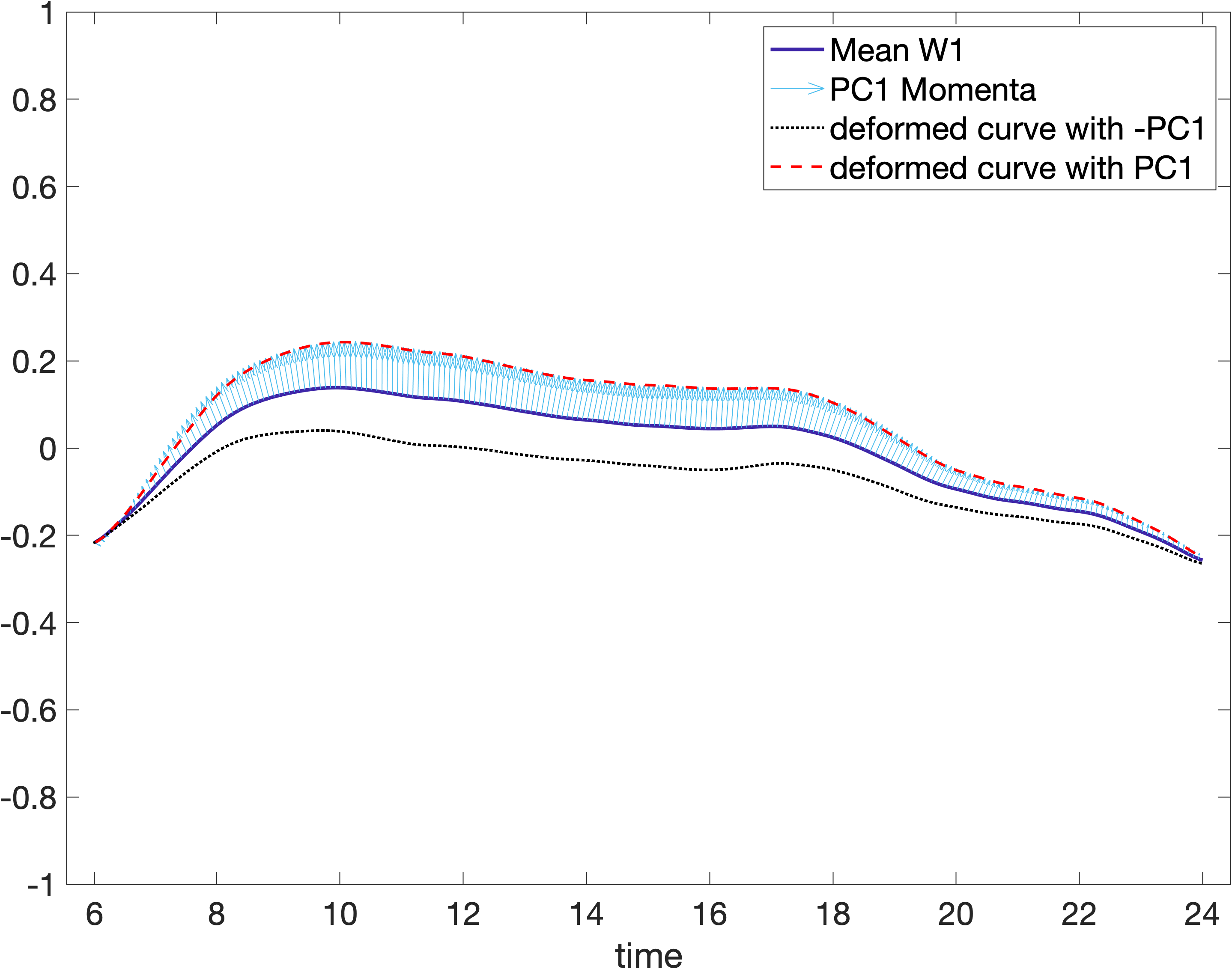}
    \captionsetup{justification=centering} 
    \caption{PC1 initial momenta and deformations in period W1--W2}
    \label{subfig:pc1_V1V2}
  \end{subfigure}
    \caption{MFPCA PC1 eigenfunctions and deformations}\label{fig:MFPCA_efunc1}
\end{figure}

\begin{figure}[h]
  \centering

    \begin{subfigure}[b]{0.48\textwidth}
    \centering
    \includegraphics[width=\textwidth]{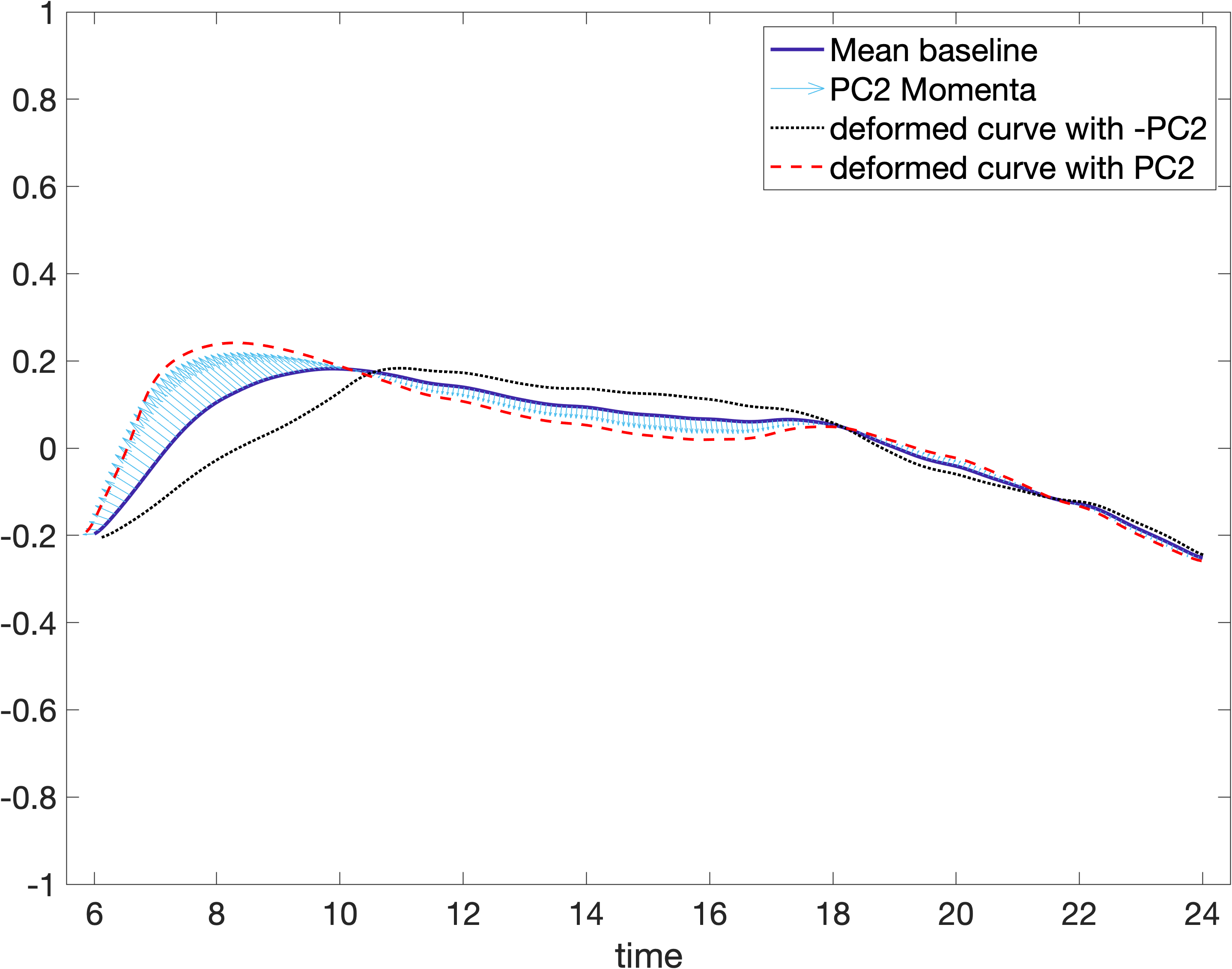}
    \captionsetup{justification=centering} 
    \caption{PC2 initial momenta and deformations in period baseline--W1}
    \label{subfig:pc2_V0V1_deform}
  \end{subfigure}
  \begin{subfigure}[b]{0.48\textwidth}
    \centering
    \includegraphics[width=\textwidth]{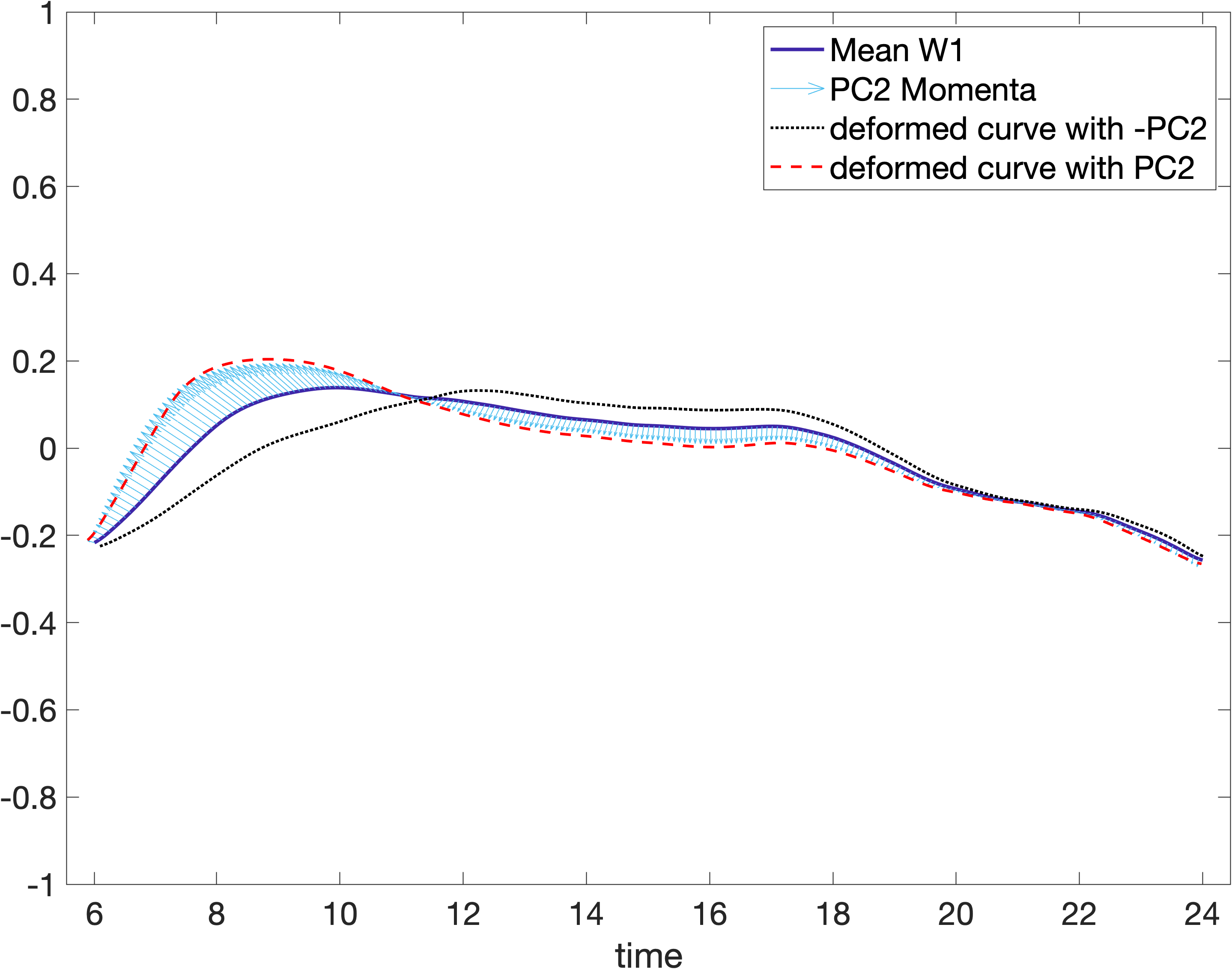}
    \captionsetup{justification=centering} 
    \caption{PC2 initial momenta and deformations in period W1--W2}
    \label{subfig:pc2_V1V2_deform}
  \end{subfigure}
    \caption{MFPCA PC2 deformations}\label{fig:MFPCA_efunc2_deform}
\end{figure}

The comparisons between the MFPCA and concatenated UFPCA mean functions (top) and first PC eigenfunctions (bottom) are shown in Supplemental (\ref{MFPCAvsUFPCA}) Figure~\ref{fig:MFPCAvsUFPCA}. The two approaches yield generally similar mean structures and PC1 patterns across both domains and both periods. However, the solid curves (UFPCA) exhibit a “smooth-over effect” at the transition between the end of the $x$-domain and the beginning of the $y$-domain (solid lines can be connected between left and right panels) , a feature that is absent in the dashed curves (MFPCA). This occurs because concatenation forced the two domains to join smoothly, thereby imposing continuity at the boundary and inducing contamination into the data structure between two domains.

Additional comparisons for PC2-PC5 are provided in Supplemental (\ref{MFPCAvsUFPCA}) Figure~\ref{fig:mfpca_ufpca_pc}. Overall, the two methods produced consistent patterns, particularly for the four leading PCs. This consistency is likely a benefit from the centering and scaling applied to the $x$- and $y$-domain functions prior to decomposition, which place the two domains on comparable scales and reduced the chance that concatenation would artificially overweight one domain or obscure domain-specific structure.

\subsection{Associations between longitudinal PA change and PF}
For association analyses, two LMMs were described in Section (\ref{method::association}) and LRT revealed significant deformation-energy-by-period interaction in both models ($p < 0.01$). In \textbf{Model 1}, LASSO removed PC2, PC3, PC4, baseline smoking status, and DBP; in \textbf{Model 2}, baseline smoking status and DBP were removed as well.

Because PF and accelerometer assessments were not collected concurrently, PF was missing for 260 participants at W1 and 338 at W2 even after allowing a $\pm$190-day matching window. Several baseline covariates also had substantial missingness (e.g., 423 missing Reynolds Risk Score and 329 missing cholesterol/glucose/blood pressure measures). Furthermore, 106 participants with previous CVD diagnosis at baseline were excluded. After restricting to participants with complete data on PF and all retained covariates, the analytic sample size for both \textbf{Model 1} and \textbf{Model 2} was $N = 1{,}157$. Significant terms of fixed-effect estimates are reported in Tables~(\ref{tab:lmm_results}).

In \textbf{Model 1}, PC1 is positively associated with PF score ($\beta$ = 60.1, s.e = 14.51, $p<0.0001$) after adjusting for Lasso-selected covariates including baseline PF score, where increase in PC1 corresponding to an upward deformation as shown in Figures 4c, 4d. That is, increasing the PC1 score from 25$^{th}$ to 75$^{th}$ percentile, which corresponds to a 0.02-unit increase, is associated with an increase of 1.3-units in the PF score (s.e = 0.33).

\begin{figure}[h]
  \centering
 \begin{subfigure}[b]{0.48\textwidth}
    \centering
    \includegraphics[width=\textwidth]{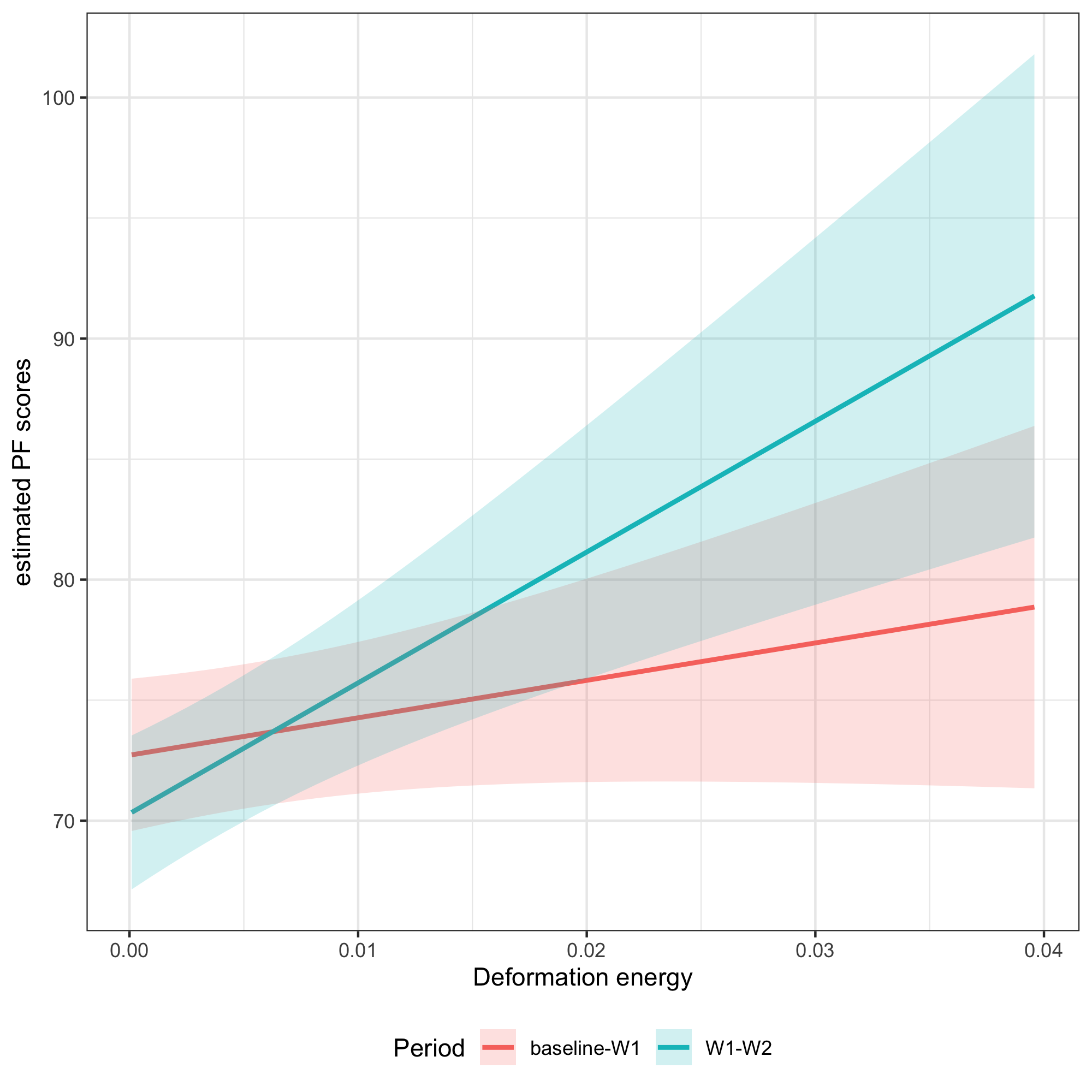}
    \captionsetup{justification=centering} 
    \caption{Model 1}\label{subfig:interaction1}
  \end{subfigure}
  \hspace{0.02\textwidth} 
  \begin{subfigure}[b]{0.48\textwidth}
    \centering
    \includegraphics[width=\textwidth]{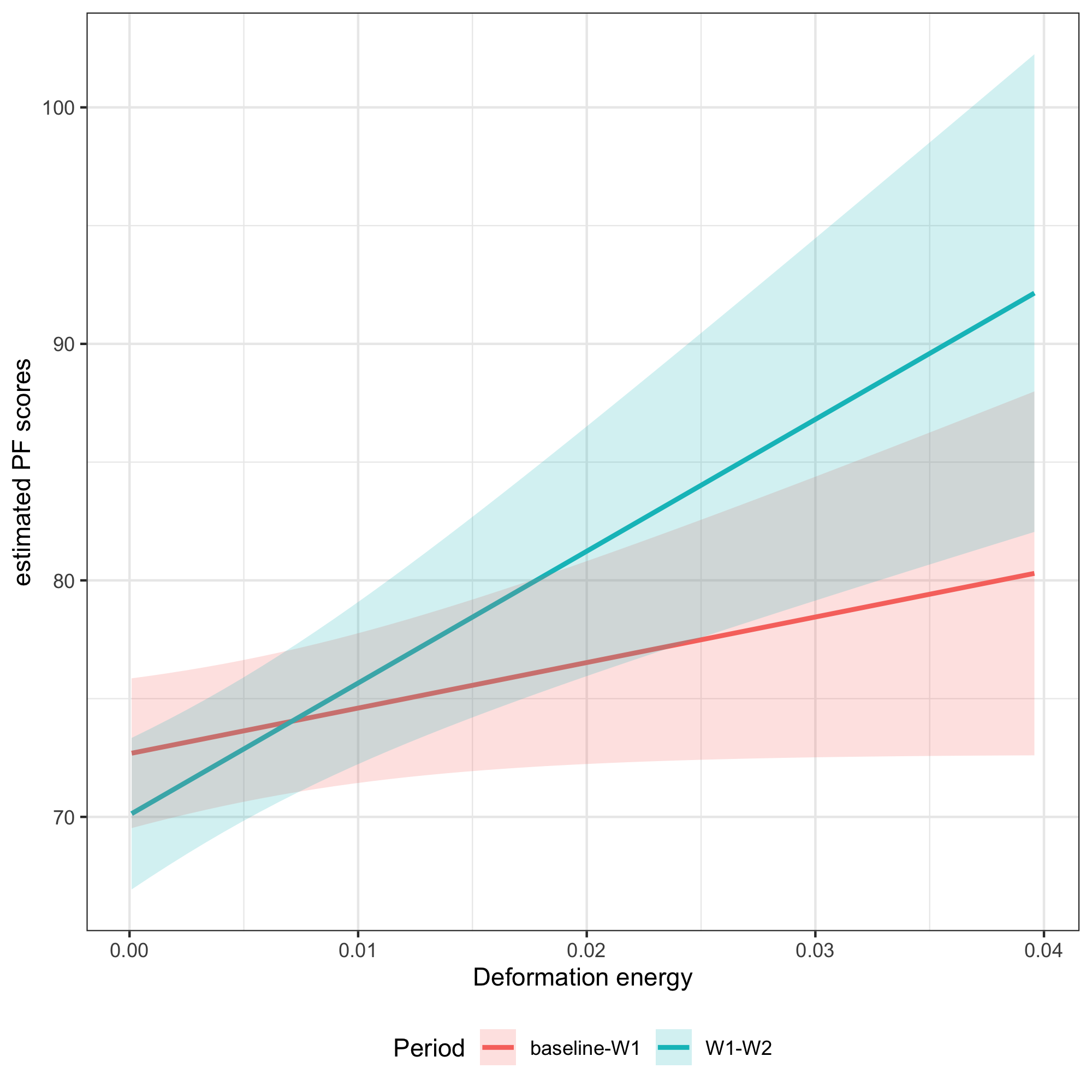}
    \captionsetup{justification=centering} 
      \caption{Model 2}\label{subfig:interaction2}
  \end{subfigure}
  \caption{Deformation energy by period interaction}\label{fig:interaction}
\end{figure}

Figure~\ref{subfig:interaction1} visualizes the interaction over the observed deformation-energy range [0, 0.04]. The deformation-energy-per-period interaction is significant ($\beta$ = 387.4, s.e = 129.84, $p = 0.003$), indicating that an increase of 0.04 in deformation energy has a 15.5 greater effect on PF in W1--W2 than in baseline--W1. This pattern suggests that PF may have been more sensitive to overall PA-pattern change in the later period, possibly because participants were older during W1--W2 and may have accumulated more prolonged sedentary time, making PF more responsive even to modest perturbations in PA patterns. 

Among baseline covariates, age and BMI are negatively associated with outcome: one year increase in age is associated with 0.5 unit decrease in PF score (s.e = 0.10, $p < 0.0001$) whereas 1 kg/m$^2$ increase in BMI is associated with 0.3 unit decrease in PF score (s.e = 0.09, $p = 0.0004$). Self-rated good (vs excellent) health is associated with lower PF (2.4 units, s.e = 0.99, $p=0.02$). Higher baseline MVPA is associated with higher PF: one hour increase in baseline daily MVPA time is associated with 1.8 unit higher in PF score (s.e = 0.6, $p=0.02$) after fixing other predictors and covariates. 

Results are similar in \textbf{Model 2}.  
$\Delta$net-AUC, used as a scalar summary of total activity change between consecutive visits, is positively associated with PF: a one-unit increase in $\Delta$net-AUC is associated with a 0.01 unit increase in PF score (s.e = 0.003, $p =0.0001$). Note that $\Delta$net-AUC was derived based on scaled net-AUC values, a one-unit change in $\Delta$net-AUC is small relative to scale of PF (0 -- 100). The deformation-energy-per-period interaction remains significant ($p=0.005$), indicating an increase of 0.04 over observed deformation energy range has a 14.6 greater effect on PF in W1--W2 than in baseline--W1 (Figure~\ref{subfig:interaction2}). Associations between significant baseline characteristics and PF in \textbf{Model 2} are similar to those from \textbf{Model 1}.

\begin{table}[ht]
\centering
\caption{Association results from 2 LMMs (N = $1,157$)}\label{tab:lmm_results}
\begin{tabular}{llcc}
\hline
& \textbf{Variable} & \textbf{Coefficient (s.e$^1$)} & \textbf{p-value} \\
\hline
\multirow{3}{*}{Model 1}\\
&Age (year)                & -0.5 (0.10) & $<$ 0.0001  \\
&BMI (kg/m$^2$)            & -0.3 (0.09) & 0.0004 \\
&MVPA (minutes/day)            &0.03 (0.01)& 0.02 \\
&RAND-36 PF Score 		 & 0.7 (0.02)& $<$ 0.0001 \\
&Self-rated health (ref: Excellent)           &               &       \\
&\quad Good           	& -2.4 (0.99) & 0.02 \\
&\quad Poor          	& -3.4 (2.61) & 0.19 \\\\

&PC1    				& 60.1(14.51) 		& $<$ 0.0001 \\
&Deformation energy    				& 155.2 (102.02) 	& 0.13 \\
&Period (ref: baseline--W1)	&               		&       \\
&\quad W1--W2			& -2.4 (0.74)  		& 0.001 \\
&Deformation energy$\times$Period	& 387.4 (129.84) 	& 0.003 \\\\

\hline
\multirow{3}{*}{Model 2}\\
&Age (year)                    & -0.5 (0.10) & $<$ 0.0001  \\
&BMI (kg/m$^2$)            & -0.3 (0.09) & 0.0004 \\
&MVPA (minutes/day)            &0.03 (0.01)& 0.02 \\
&RAND-36 PF Score 		 & 0.7 (0.02)& $<$ 0.0001 \\
&Self-rated health (ref: Excellent)           &               &       \\
&\quad Good           & -2.4 (0.99) & 0.02 \\
&\quad Poor          & -3.4 (2.61) & 0.19 \\\\

&$\Delta$net-AUC			& 0.01 (0.003)		& 0.0001 \\
&Deformation energy    				& 192.5 (104.50) 	& 0.07 \\
&Period (ref: baseline--W1)	&               		&       \\
&\quad W1--W2      		& -2.6 (0.74) 		& 0.0005 \\
&Deformation energy$\times$Period    	& 364.9 (129.57) 	& 0.005 \\\\
\hline
\multicolumn{3}{l}{\footnotesize $^1$ standard error} \\
\end{tabular}
\end{table}


\section{Conclusion}\label{sec:discussion}
Prior OPACH analyses have established associations between accelerometer-measured PA and mortality \citep{LaMonte:2018jb, nguyen2024prospective}, CVD and heart failure \citep{lamonte2024accelerometer, Bellettiere2019a}, cancer \citep{hyde2025sitting}, and blood glucose \citep{wang2023robust}. A longitudinal OPACH--WHISH study reported significant PA decline over 7 years, with the largest decline observed for MVPA \citep{hyde2025aging}. Recent work from the Women's Health Accelerometry Collaboration (WHAC) showed that higher accelerometer-measured PA was associated with improved odds of survival to age 90 with and without mobility disability \citep{hyde2025prospective, lee2018using}. Cross-sectional evidence further supports associations between sensor-based PA and physical fitness \citep{su2025association, caspersen1985physical}, as well as between PA and PF measured by the Short Physical Performance Battery \citep{corcoran2016accelerometer, rantalainen2022complexity}. In these previous analyses, the accelerometer output typically was processed using conventional framework resulting in averaged VM count time-series data, which might obscure important shifts in habitual PA amounts and patterns throughout the day.

Our present work extends this literature by shifting the analytic target from static PA levels to \emph{longitudinal change in diurnal PA patterns}. Rather than relying solely on scalar summaries, we modeled visit-to-visit change as Riemannian diffeomorphic deformations between diurnal PA curves and used MFPCA to obtain clinically interpretable change phenotypes. After removal of the mean function, PC1 represented the residual dominant mode of diurnal PA change (increase/decline) across the day, especially between 10 a.m. and 7 p.m.. It was strongly associated with PF after adjustment for baseline PF and covariates, supporting the clinical relevance of deformation-derived change features. We also observed a positive association between PF and $\Delta$net-AUC, a signed scalar measure of total activity change, but $\Delta$net-AUC does not distinguish whether change is driven by timing reallocation, magnitude shifts at specific times of day, or both. Deformation energy provided a complementary scalar index of overall PA change, and its association with PF was stronger in W1--W2 than in baseline--W1, indicating greater sensitivity of later-life function to recent changes in daily activity routines. Together, these findings underscore that \emph{how} activity changes across the day carries information relevant to functional aging that is not captured by standard aggregate metrics. 

There were several limitations. First, OPACH--WHISH participants are older women; results may not generalize to other populations. Second, missing data in the LMMs may have reduced statistical power and could introduce some degree of bias in parameter estimates. Third, the randomization status was unavailable at the time of analysis; incorporating a group indicator to assess whether PF changes differed between intervention arms is an important future direction. Fourth, PF was measured by self-reported RAND-36 rather than performance-based metrics (e.g., gait speed); however, in the national WHI cohort, within which the OPACH-WHISH subset resides, RAND-36 is highly correlated with performance-based PF measures and shows similar discrimination of mortality outcomes \citep{laddu2022physical}.
Fifth, we pooled weekdays and weekends when constructing each participant's diurnal PA curve, which may mask day-type-specific patterns of change; separately modeling weekday and weekend deformations is a natural future direction of this work. 
Finally, MFPCA was applied separately to each period rather than as a single multi-level MFPCA. Although multi-level models offer advantages such as borrowing strength across periods and separating between- and within-subject components, existing frameworks are largely limited to small sample sizes due to computational challenges \citep{volkmann2023multivariate, morris2017comparison}. An important future direction is to develop scalable multi-level MFPCA method that is applicable to large cohorts like OPACH-WHISH.


\backmatter

\section{Acknowledgments}
This work used resources available through the National Research Platform (NRP) at the University of California, San Diego. NRP has been developed, and is supported in part, by funding from National Science Foundation, from awards 1730158, 1540112, 1541349, 1826967, 2112167, 2100237, and 2120019, as well as additional funding from community partners.

The authors thank WHI participants, staff, and investigators. 
Program Office: (National Heart, Lung, and Blood Institute, Bethesda, Maryland)
Jacques Rossouw, Jared Reis, and Candice Price 
Clinical Coordinating Center: (Fred Hutchinson Cancer Center, Seattle, WA) Garnet
Anderson, Ross Prentice, Andrea LaCroix, and Charles Kooperberg
Steering Committee and Academic Centers: (University of California, Davis) Lorena Garcia; (Wake Forest University) Lindsay Reynolds; (University at Buffalo) Amy Millen; (University at Buffalo) Jean Wactawski-Wende; (Fred Hutchinson Cancer Center) Marian Neuhouser; (Fred Hutchinson Cancer Center) Holly Harris; (University of Massachusetts) Brian Silver; (University of Tennessee Health Center)Karen Johnson; (Stanford Prevention Research Center) Marcia L. Stefanick; (The Ohio State University) Electra Paskett; (Wake Forest University) Mara Vitolins. For a list of all the investigators who have contributed to WHI science, please visit: \url{https://s3-us-west-2.amazonaws.com/www-whi-org/wp-content/uploads/WHI-Investigator-Long-List.pdf}	

The OPACH Study is funded by NHLBI grants HL105065 and HL153462. The WHISH Trial is funded by NIH/NHLBI U01 HL122280-CCC, U01 HL122273-DCC, and R61/R33HL151885 (ClinicalTrials.gov NCT02425345). The WHI program is funded by the NHLBI, NIH, U.S. Department of Health and Human Services through contracts 75N92021D00001, 75N92021D00002, 75N92021D00003, 75N92021D00004, and 75N92021D00005.
\section{Ethics and consent to participants}
All participants provided written informed consent to the WHI study, which was approved by the Institutional Review Board at the Fred Hutchinson Cancer Center. Participants to the WHISH study were WHI participants who were consented using a passive “opt-out” consent, approved by the Institutional Review Board at the Fred Hutchinson Cancer Center. University of California at San Diego exempted Institutional Review Board approval for the use of existing data for the present study.


\section{Conflicts of Interest and Source of Funding}
The authors declare no potential conflict of interests. JZ's, LD's, and AL's work was supported by NIH/NHLBI R01HL166802. 
RZ's and LN's work was supported by NIH/NHLBI R01HL166802 and NIH/NHLBI 1R01HL168535. 
CD's work was supported by NIH/NHLBI R01HL130483. RZ's and LN's work was partially supported by NIH/NHLBI R01HL130483.

\section{Availability of data and materials}
The authors do not have permission to share data.



\clearpage
\section{Supplemental Digital Content}\label{supp}
\subsection{Multivariate FPCA Estimation}\label{MFPCA_steps}

The univariate truncated Karhunen--Lo\`eve expansion of FPCA can be expressed as
\begin{equation}
S_i^{(j)}(t)
= \mu^{(j)}(t)
+ \sum_{k=1}^{K^{(j)}} \xi^{(j)}_{i,k}\, \boldsymbol{\phi}^{(j)}_{k}(t),
\qquad t \in \mathcal{T}_j.
\label{eq:univ_fpca}
\end{equation}
where $i$ is the index for participant, $i=1,2,\dots, n$;  $j$ is the index for domain and can be omitted with one domain only data; $\mu^{(j)}(t)$ represents mean function, $\{\phi^{(j)}_k\}_{k=1}^{K^{(j)}}$ and $\{\xi^{(j)}_{i,k}\}_{k=1}^{K^{(j)}}$ represent eigenfunctions and PC scores, respectively. The MFPCA consist of following steps:
\begin{enumerate}
 	\item Univariate functional decomposition on estimated initial momenta at $x$-domain ($j=1$) and $y$-domain ($j=2$), separately; default proportion of variance explained (PVE) to 99\% attaining $K^{(j)}$ components from $j^{th}$ domain and total $K =  \sum_{j=1}^2K^{(j)}$ components from both domains. 
	\item Define the matrix $\boldsymbol{\Xi}_{n\times K}$, where $n$ is the samples size (e.g, $N_1$ for basleine-W1 and $N_2$ for W1--W2). Each row, $(\xi^{(1)}_{i,1}, \dots, \xi^{(1)}_{i,K^{(1)}}, \xi^{(2)}_{i,1}, \dots, \xi^{(2)}_{i,K^{(2)}})$, contains all univariate PC scores from participant $i$ from both domains.
	\item\label{eigenanalysis} Let $\hat{\boldsymbol{Z}}_{K\times K}= (n-1)^{-1}\boldsymbol{\Xi}^T\boldsymbol{\Xi}$, and perform matrix eigen-analysis for $\hat{\boldsymbol{Z}}$ resulting in eigenvalues $\hat{\boldsymbol{\nu}}$ with length $K$ and eigenvectors $\hat{\boldsymbol{c}}_{K\times K} = $ \\
	$\begin{bmatrix} 
			\hat{\boldsymbol{c}}^{(1)}_{K^{(1)}\times K} \\\\
			\hat{\boldsymbol{c}}^{(2)}_{K^{(2)}\times K} 
	\end{bmatrix}$
	\item Select number of multivariate components, $L \le K$, based on $\hat{\boldsymbol{\nu}}$. In our application, select $L = \mbox{min} \left\{l: \frac{\nu_1 + \nu_2 + \dots + \nu_L}{\nu_1 + \nu_2 + \dots + \nu_K}  \ge 0.9 \right\}$;
	 hence, $(\nu_1, \nu_2, \dots, \nu_L)$ is also interpreted as the multivariate PC eigenvalues.
	 \item $l^{th}$ multivariate eigenfunction from $j^{th}$ domain is estimated as the linear combination of their univariate eigenfunctions using eigenvector from step \ref{eigenanalysis} as weight , 
	 \begin{equation}
	 	\hat{\boldsymbol{\psi}}^{(j)}_l(t) = \sum_{k=1}^{K^{(j)}}\hat{c}_{k,l}^{(j)}\hat{\boldsymbol{\phi}}_k^{(j)}(t)
	\end{equation}
	 \item $l^{th}$ multivariate PC score is also calculated as the linear combination of their univariate PC scores with eigenvector as weight and aggregated across both domains,
	 	 \begin{equation}
	 	\hat{\rho}_{i,l} = \sum_{j=1}^2\sum_{k=1}^{K^{(j)}}\hat{c}_{k,l}^{(j)}\hat{\xi}_{i,k}^{(j)}.
	\end{equation}
\end{enumerate}
More details are described in \citep{happ2018multivariate,MFPCA:Clara}. 

\clearpage
\subsection{PVE (\%) by PCs}\label{varExp}
\begin{table}[h]
\begin{tabular}{lcc}
\hline
& \textbf{baseline--W1} & \textbf{W1--W2} \\
\hline
PC1		 & 22.4& 20.8\\
PC2 		&  10.9 & 11.5      \\
PC3		& 9.1& 8.3 \\
PC4		 & 6.9& 7.3\\
PC5		&  6.8 & 7.1     \\
PC6		&6.2& 6.2\\
PC7		 & 5.5& 6.1\\
PC8 		& 5.2 & 5.5       \\
PC9		& 5.0& 5.2 \\
PC10	& 4.4 & 4.6  \\
PC11 	&  4.1  &  4.1     \\
PC12	& 4.0 & 3.8 \\
PC13	& 3.5 & 3.3  \\
PC14 	&  3.2 &   3.2   \\
PC15	& 3.0& 2.9\\
\hline

\end{tabular}
\end{table}

\clearpage
\subsection{Additional MFPCA PCs}\label{otherPCs}

\begin{figure}[h]
  \centering

  \begin{subfigure}[b]{0.48\textwidth}
    \centering
    \includegraphics[width=\textwidth]{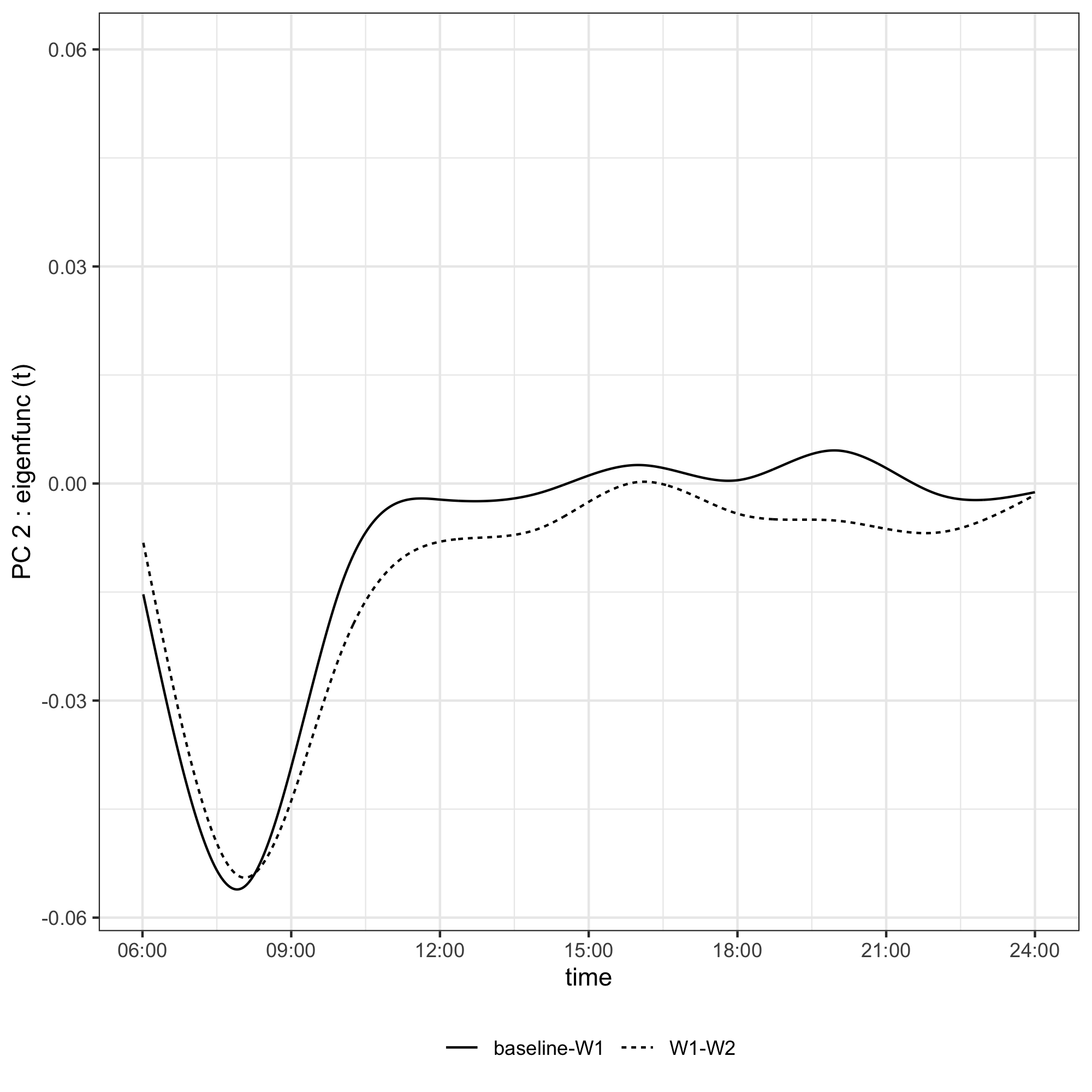}
     \captionsetup{justification=centering} 
    \caption{PC2 eigenfunction at x domain }
    \label{subfig:pc2X}
  \end{subfigure}
  \begin{subfigure}[b]{0.48\textwidth}
    \centering
    \includegraphics[width=\textwidth]{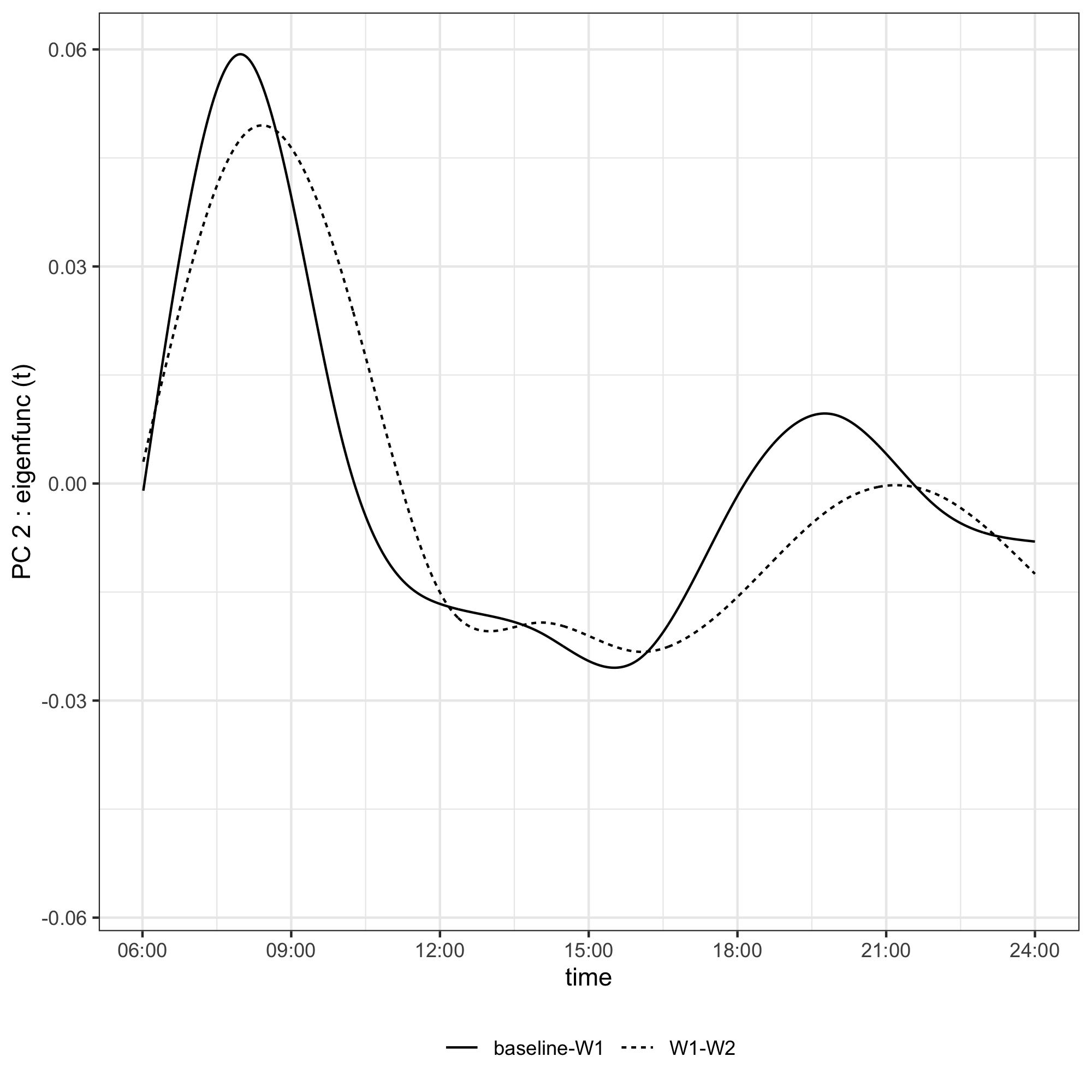}
    \captionsetup{justification=centering} 
    \caption{PC2 eigenfunction at y domain}
    \label{subfig:pc2Y}
  \end{subfigure}

  \vspace{0.8em}

    \begin{subfigure}[b]{0.48\textwidth}
    \centering
    \includegraphics[width=\textwidth]{Figures/fpca/deformation/baseline_V1_mFPCA_M15_PC2.png}
    \captionsetup{justification=centering} 
    \caption{PC2 eigenfunction based initial momenta and deformations baseline--W1}
    \label{subfig:pc2_V0V1}
  \end{subfigure}
  \begin{subfigure}[b]{0.48\textwidth}
    \centering
    \includegraphics[width=\textwidth]{Figures/fpca/deformation/V1_V2_mFPCA_M15_PC2.png}
    \captionsetup{justification=centering} 
    \caption{PC2 eigenfunction based initial momenta and deformations W1--W2 }
    \label{subfig:pc2_V1V2}
  \end{subfigure}
    \caption{MFPCA PC2 eigenfunctions and deformations. PC2 shows consistent temporal shifts and magnitude changes across the two periods, a left shift and localized increase in the morning activity is illustrated by the up-leftward-pointing arrows in Figure \ref{subfig:pc2_V0V1} for baseline --W1 (6:00 a.m. to 10:18 a.m.) and Figure \ref{subfig:pc2_V1V2} for W1--W2 (6:00 a.m. to 11:13 a.m),  followed by a slight downward arrows.
}\label{fig:MFPCA_efunc2}
\end{figure}

\begin{figure}[h]
  \centering

  \begin{subfigure}[b]{0.48\textwidth}
    \centering
    \includegraphics[width=\textwidth]{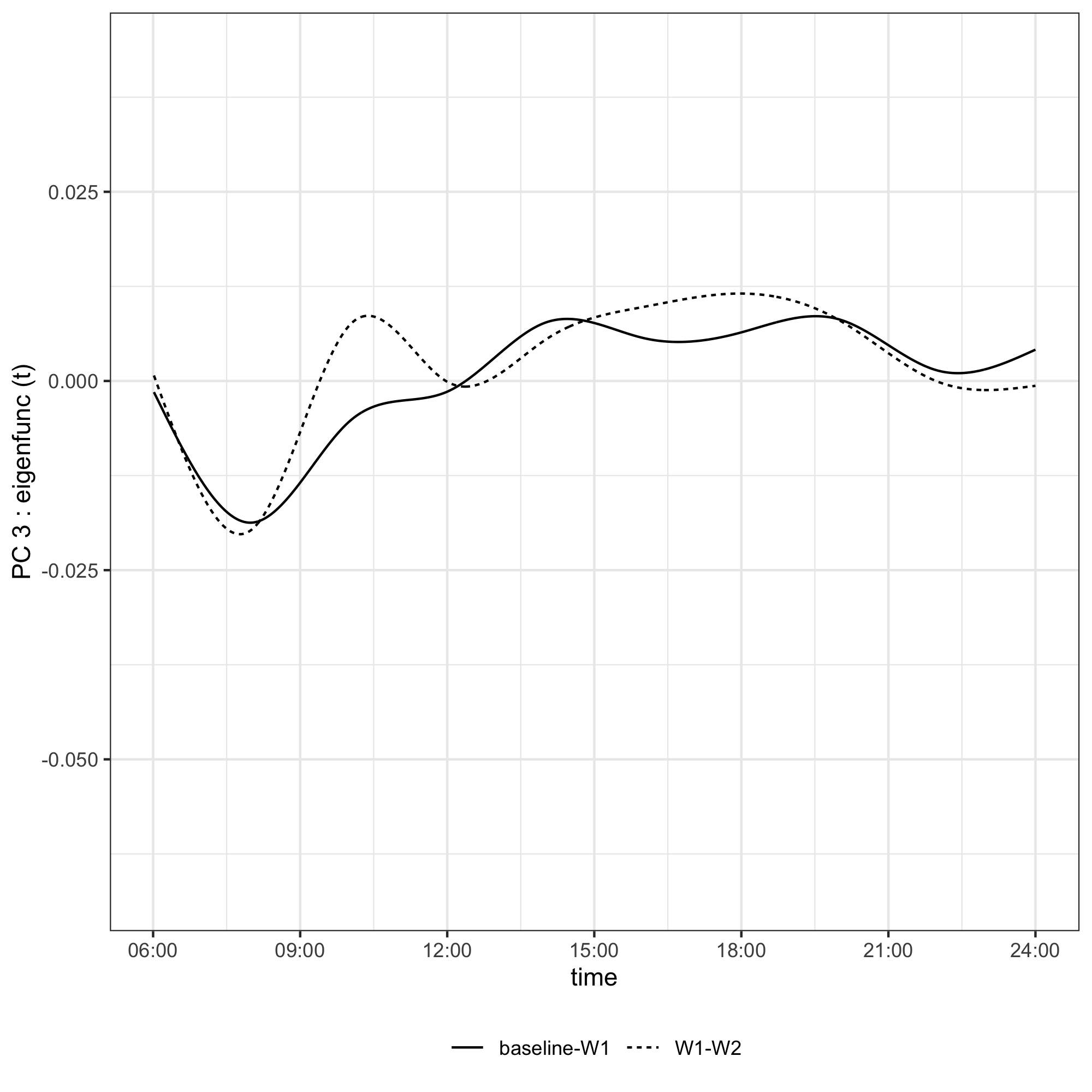}
     \captionsetup{justification=centering} 
    \caption{PC3 eigenfunction at x domain }
    \label{subfig:pc3X}
  \end{subfigure}
  \begin{subfigure}[b]{0.48\textwidth}
    \centering
    \includegraphics[width=\textwidth]{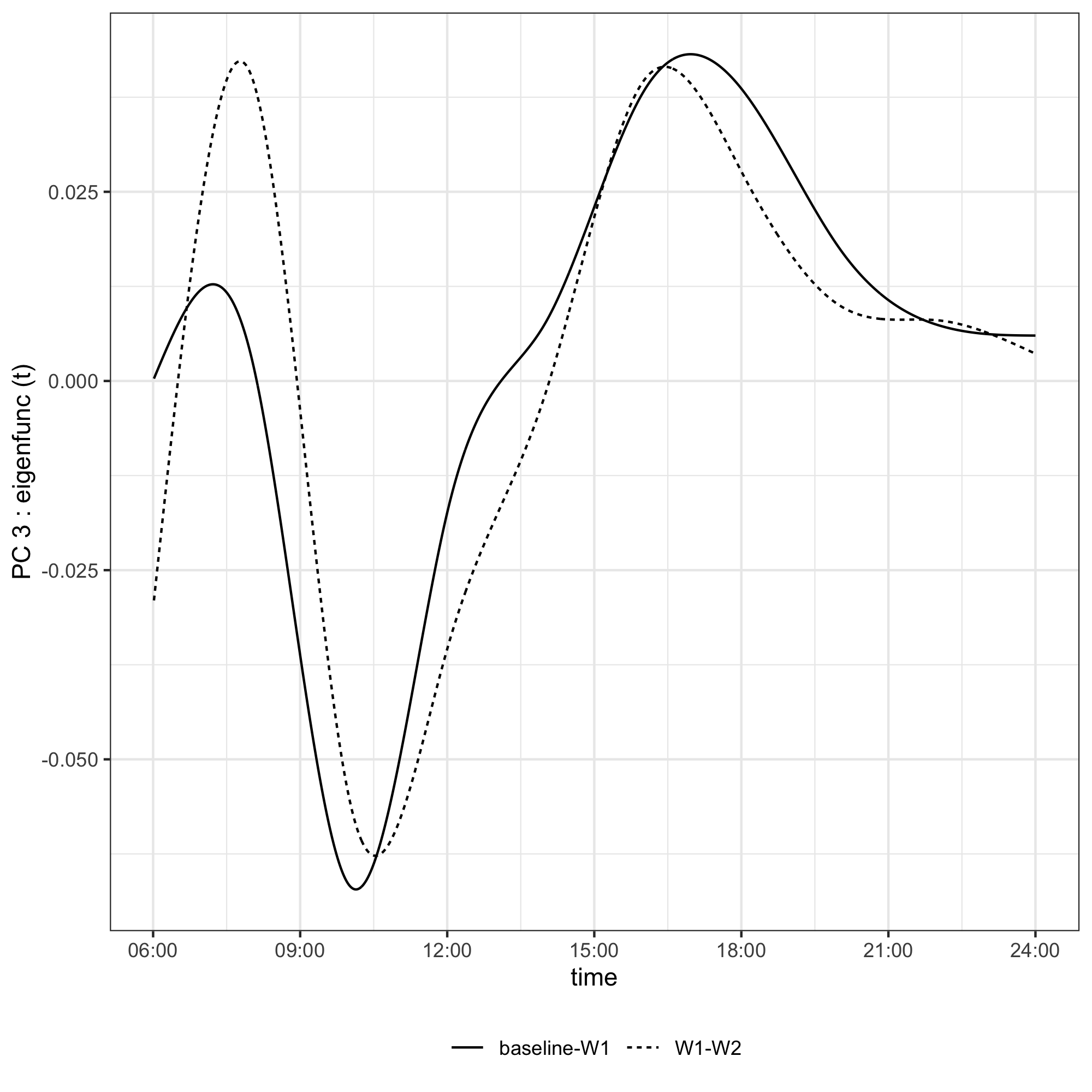}
    \captionsetup{justification=centering} 
    \caption{PC3 eigenfunction at y domain}
    \label{subfig:pc3Y}
  \end{subfigure}

 \vspace{0.8em}

    \begin{subfigure}[b]{0.48\textwidth}
    \centering
    \includegraphics[width=\textwidth]{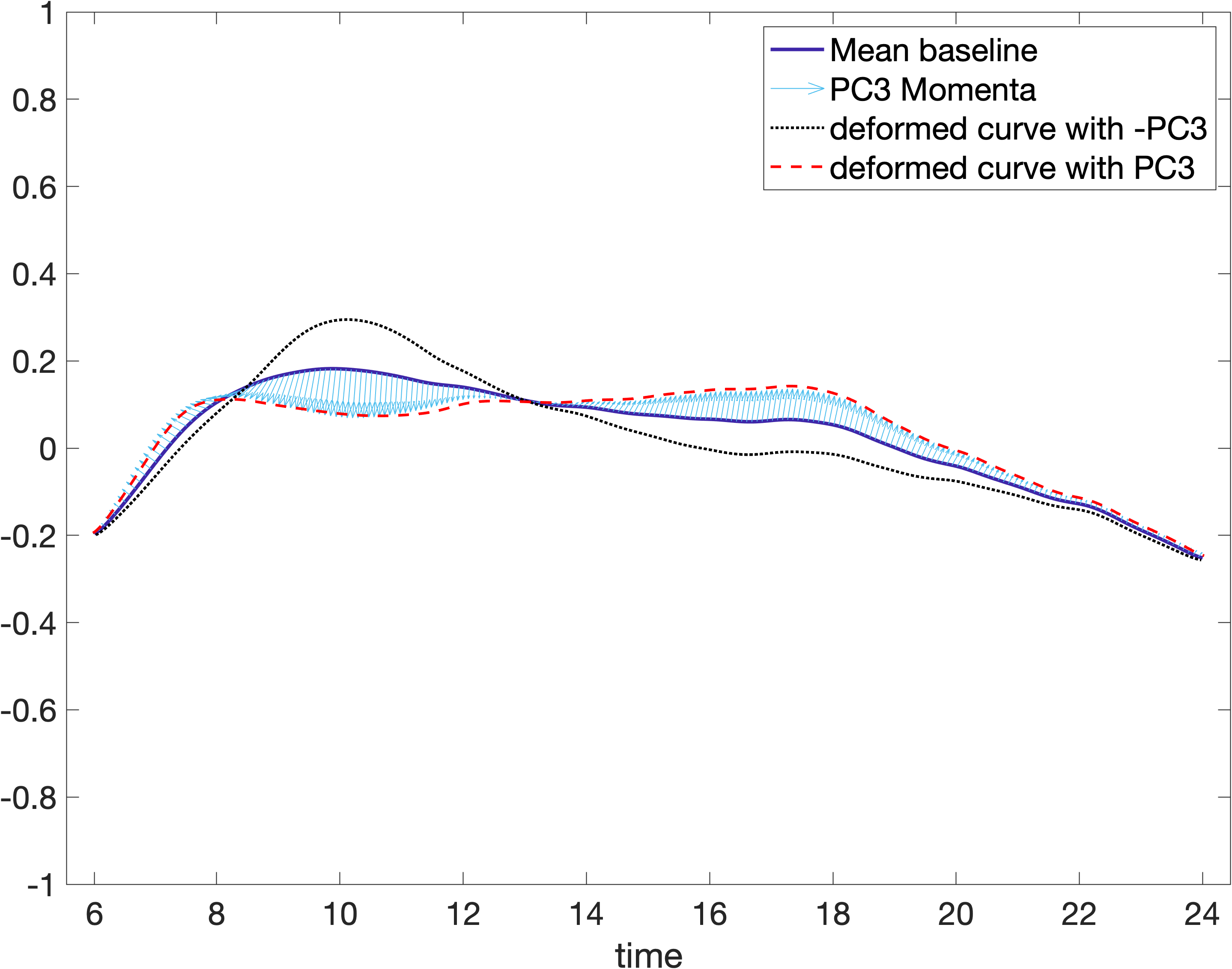}
    \captionsetup{justification=centering} 
    \caption{PC3 eigenfunction based initial momenta and deformations baseline--W1}
    \label{subfig:pc3_V0V1}
  \end{subfigure}
  \begin{subfigure}[b]{0.48\textwidth}
    \centering
    \includegraphics[width=\textwidth]{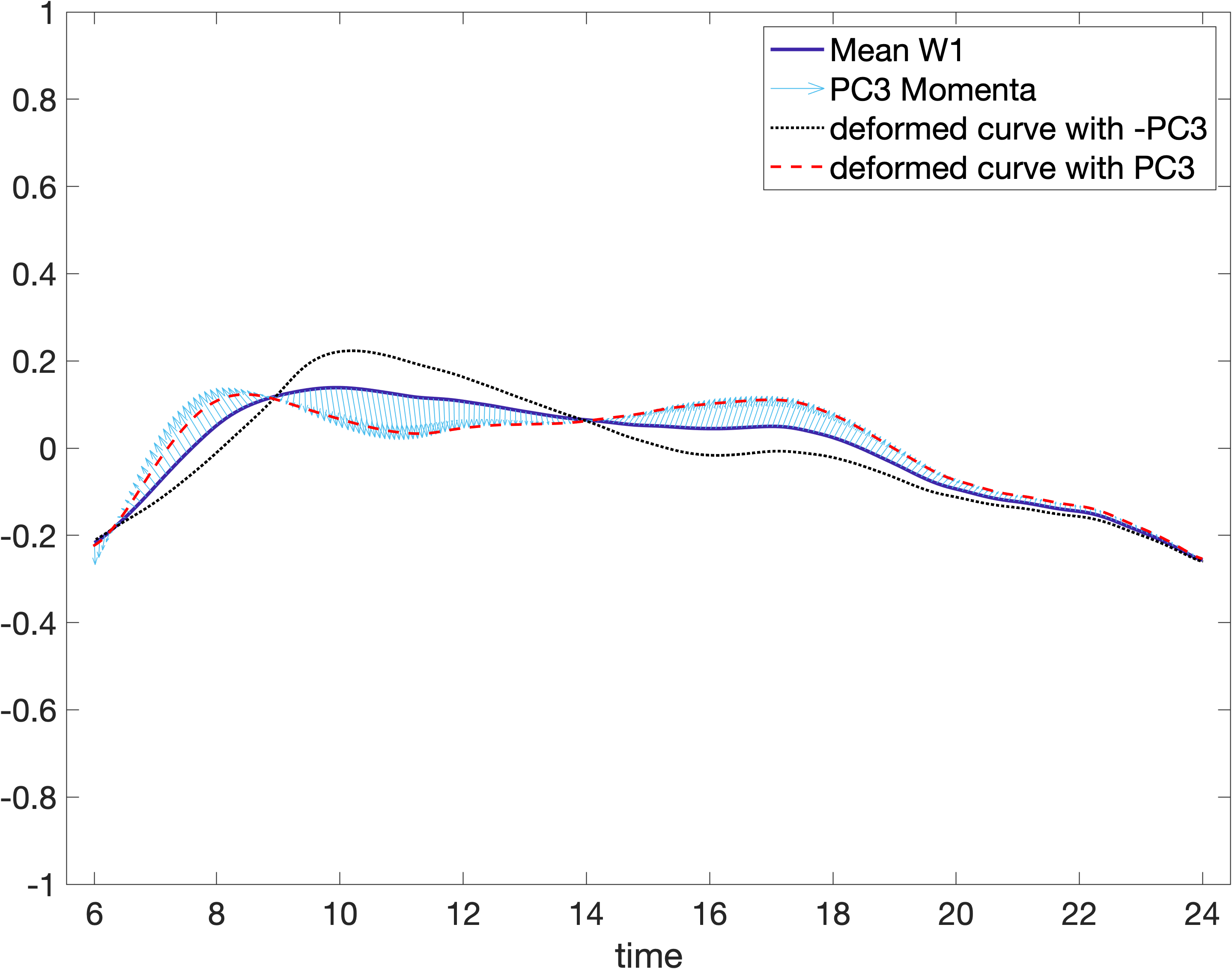}
    \captionsetup{justification=centering} 
    \caption{PC3 eigenfunction based initial momenta and deformations W1--W2 }
    \label{subfig:pc3_V1V2}
  \end{subfigure}
    \caption{MFPCA PC3 eigenfunctions and deformations. PC3 exhibits a redistribution of PA in both periods. Activity levels increase leftwards in the early morning (6 a.m to around 9 a.m.) and decrease afterwords until around 2 p.m., then increase again slightly shift towards right until end of the day.}\label{fig:MFPCA_efunc3}
\end{figure}

\begin{figure}[h]
  \centering

  \begin{subfigure}[b]{0.48\textwidth}
    \centering
    \includegraphics[width=\textwidth]{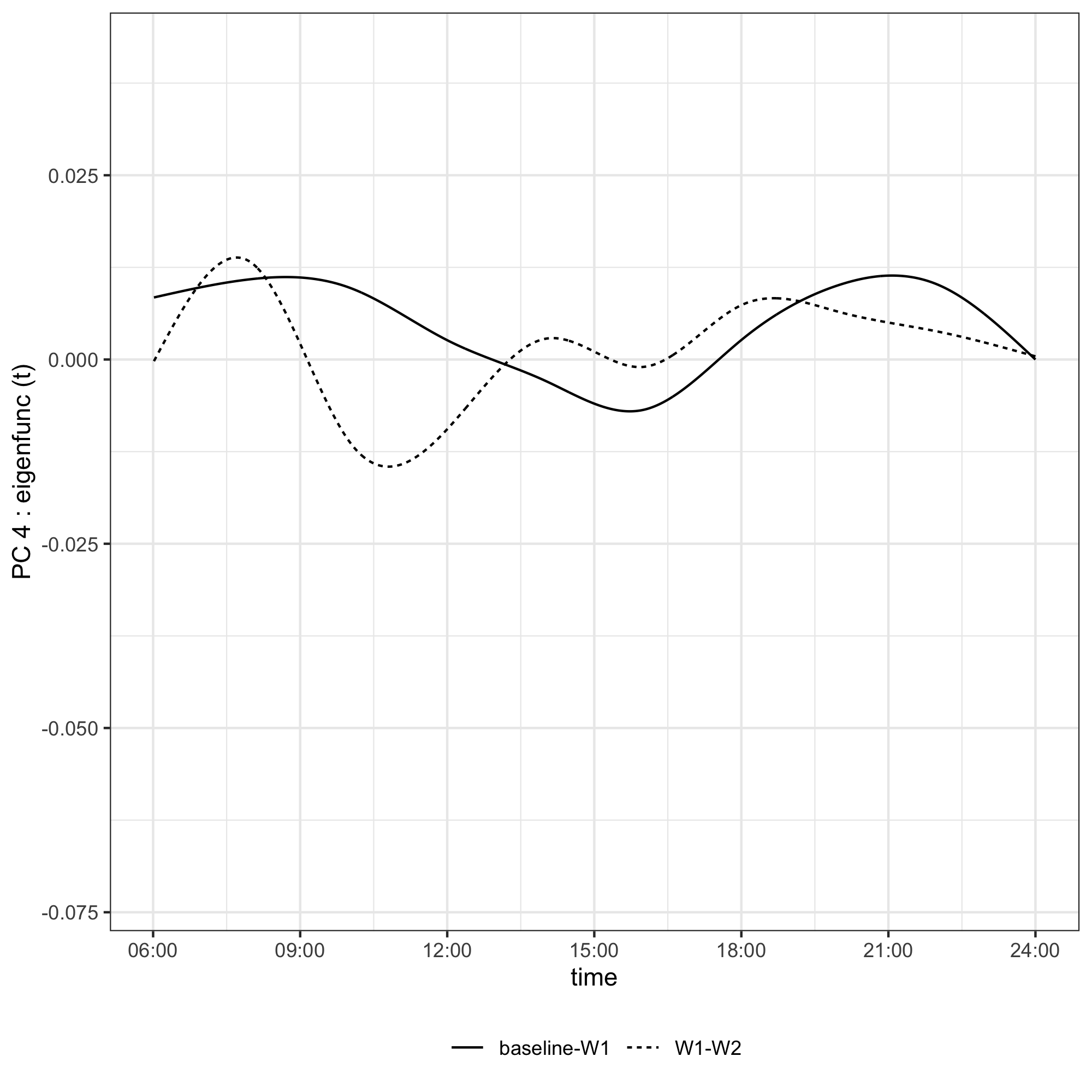}
     \captionsetup{justification=centering} 
    \caption{PC4 eigenfunction at x domain }
    \label{subfig:pc4X}
  \end{subfigure}
  \begin{subfigure}[b]{0.48\textwidth}
    \centering
    \includegraphics[width=\textwidth]{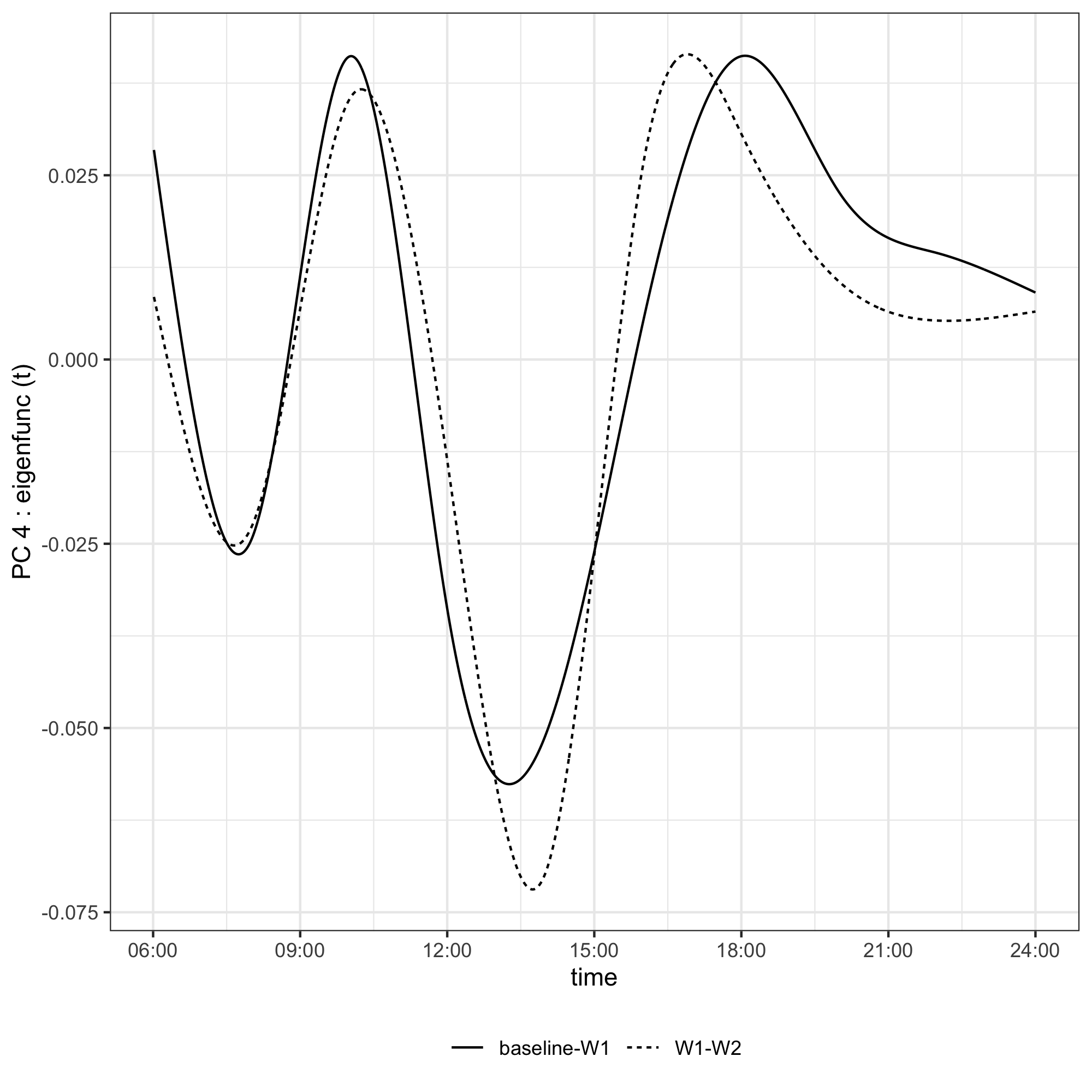}
    \captionsetup{justification=centering} 
    \caption{PC4 eigenfunction at y domain}
    \label{subfig:pc4Y}
  \end{subfigure}

  \vspace{0.8em}

    \begin{subfigure}[b]{0.48\textwidth}
    \centering
    \includegraphics[width=\textwidth]{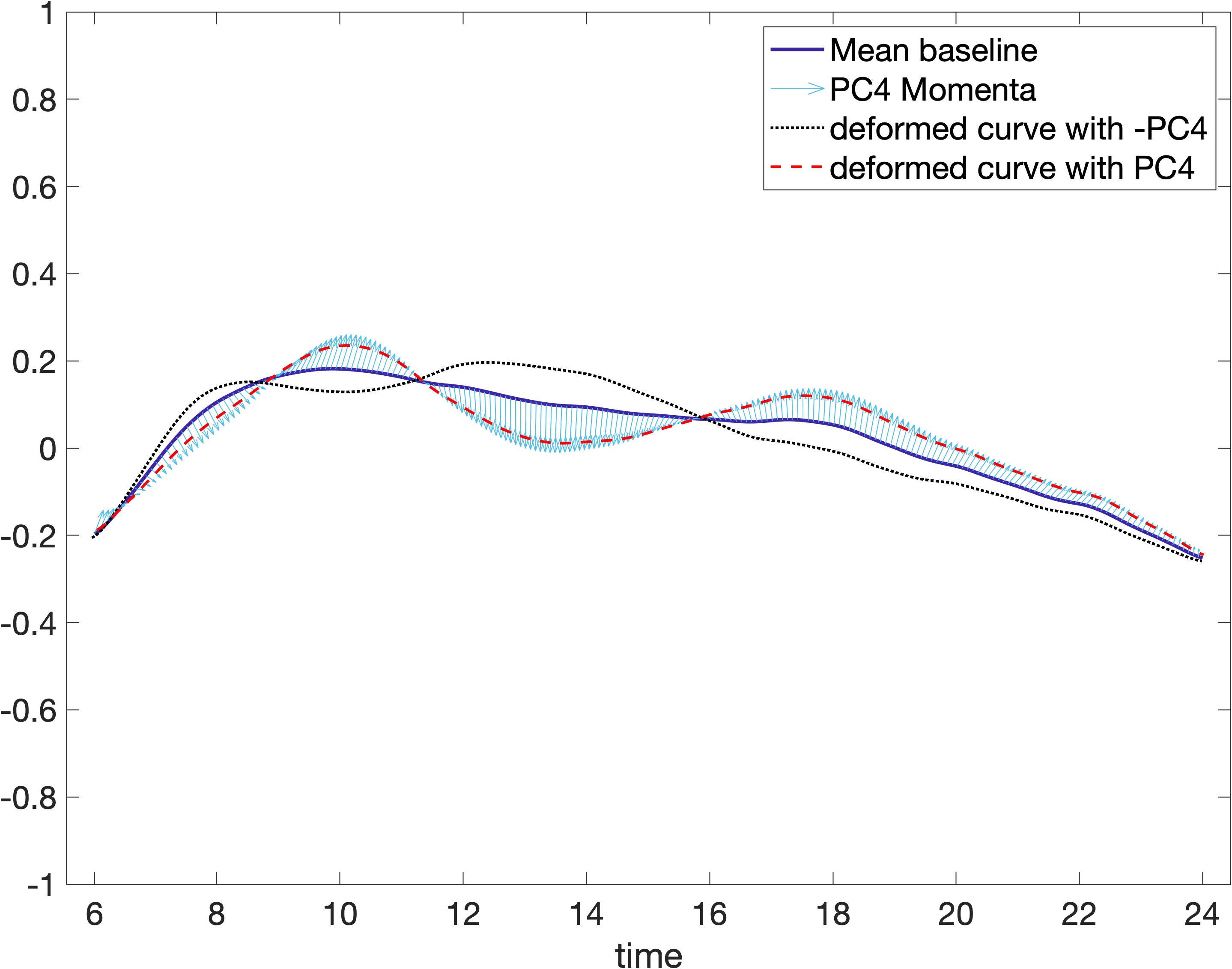}
    \captionsetup{justification=centering} 
    \caption{PC4 eigenfunction based initial momenta and deformations baseline--W1}
    \label{subfig:pc4_V0V1}
  \end{subfigure}
  \begin{subfigure}[b]{0.48\textwidth}
    \centering
    \includegraphics[width=\textwidth]{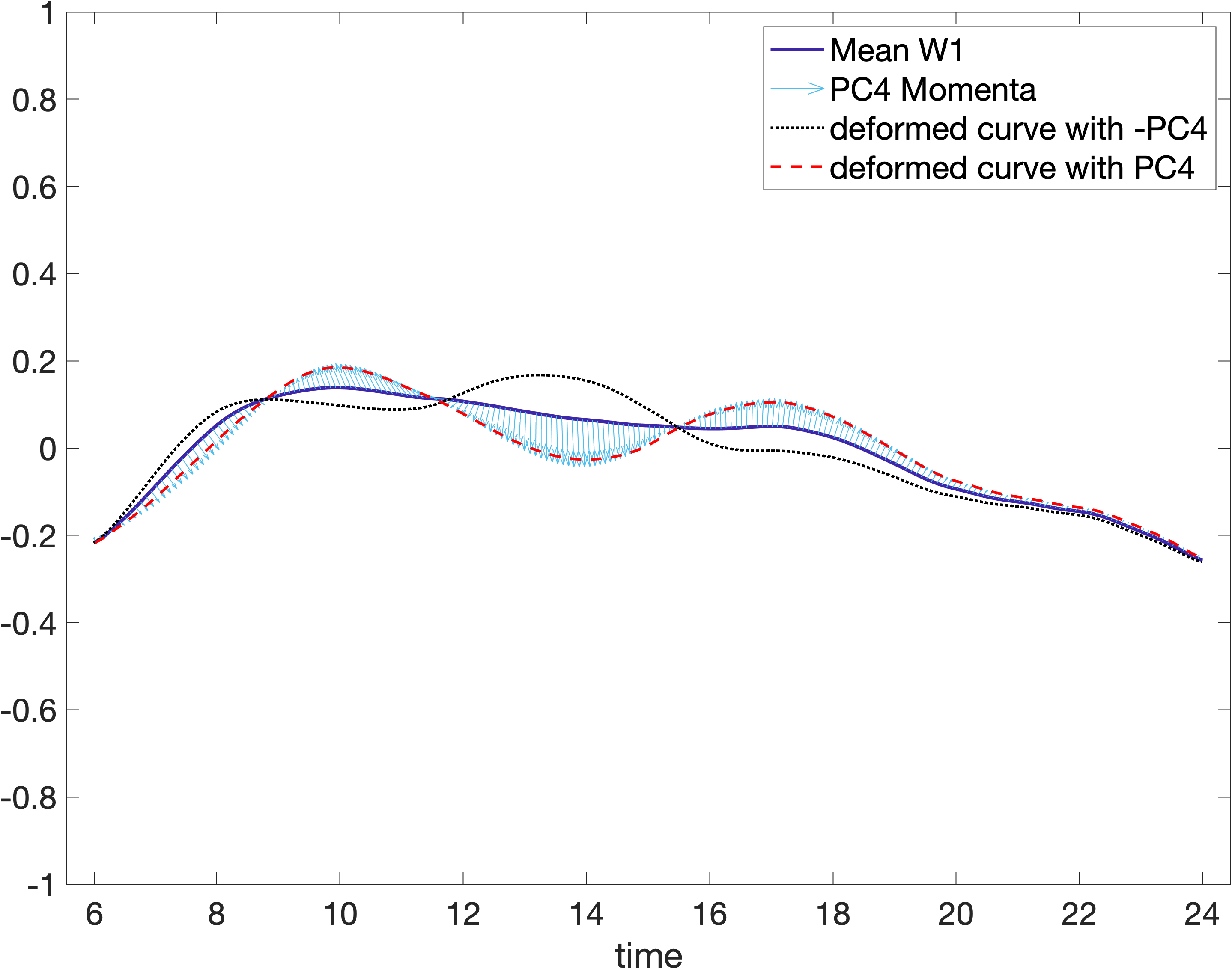}
    \captionsetup{justification=centering} 
    \caption{PC4 eigenfunction based initial momenta and deformations W1--W2}
    \label{subfig:pc4_V1V2}
  \end{subfigure}
    \caption{MFPCA PC4 eigenfunctions and deformations. PC4 also display similar pattern of PA redistribution in both periods. Activity levels decrease towards right until around 9 a.m. and increase afterwards until noon; decrease downwards again until around 4 p.m. then increase until the end of the day.}\label{fig:MFPCA_efunc4}
\end{figure}

\begin{figure}[h]
  \centering

  \begin{subfigure}[b]{0.48\textwidth}
    \centering
    \includegraphics[width=\textwidth]{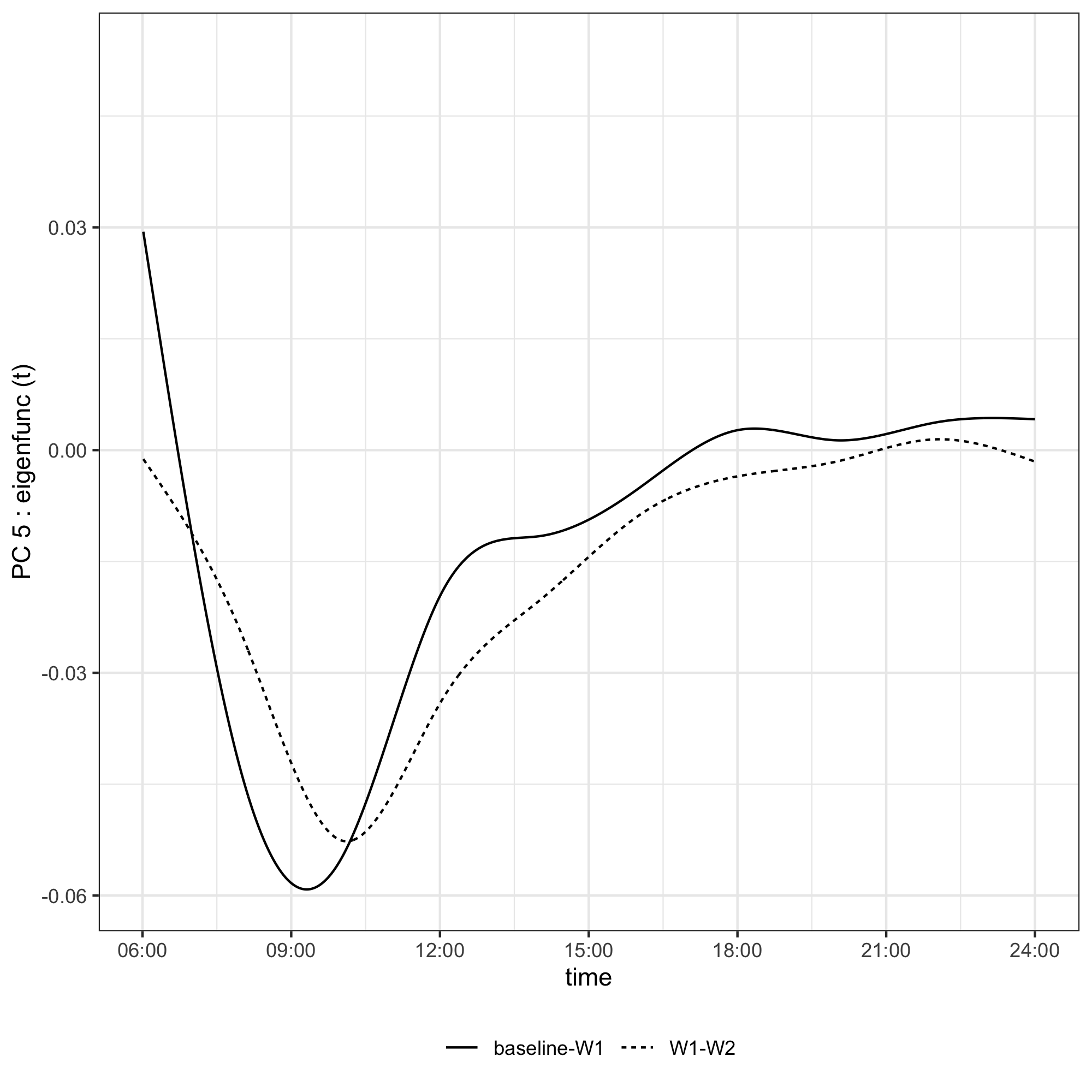}
     \captionsetup{justification=centering} 
    \caption{PC5 eigenfunction at x domain }
    \label{subfig:pc5X}
  \end{subfigure}
  \begin{subfigure}[b]{0.48\textwidth}
    \centering
    \includegraphics[width=\textwidth]{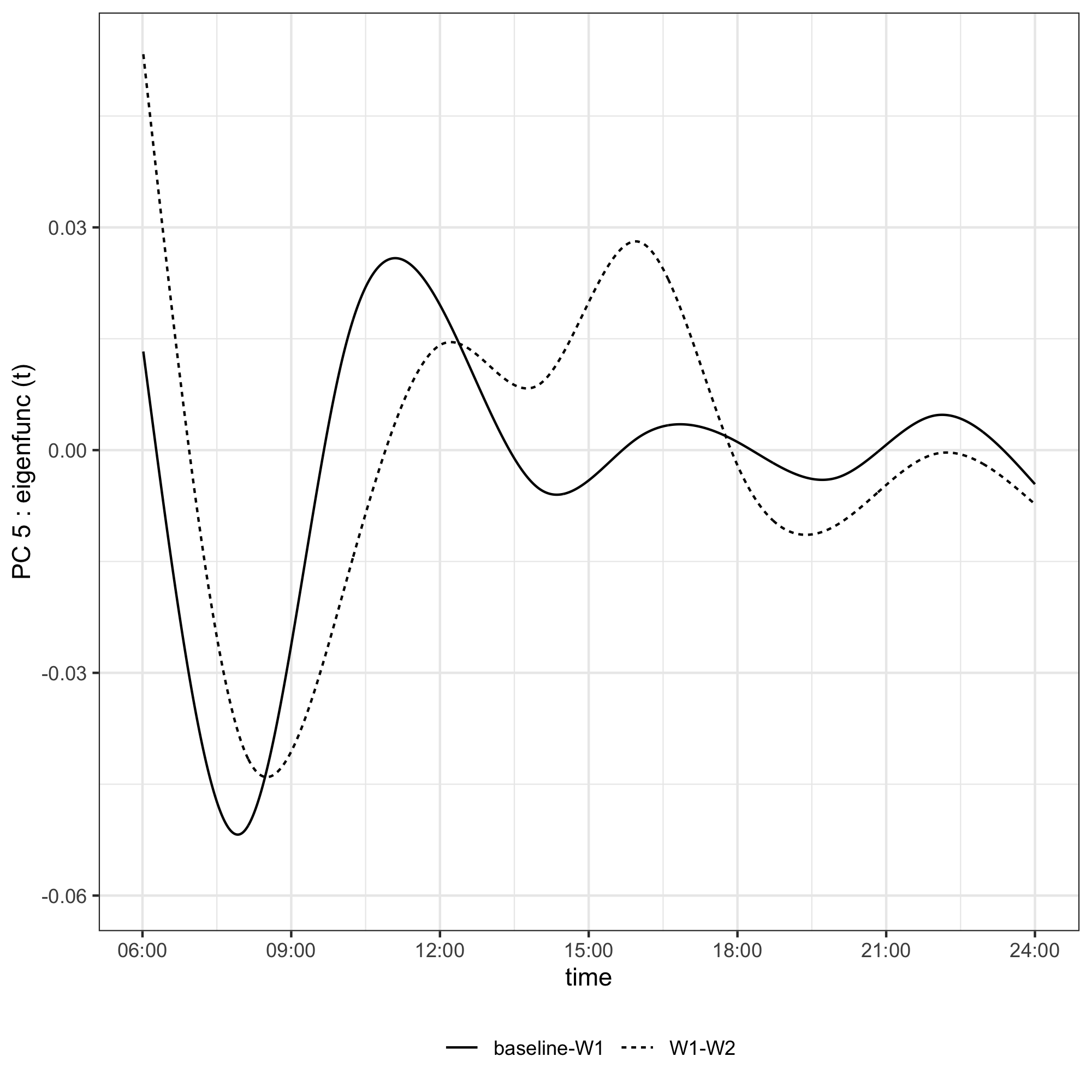}
    \captionsetup{justification=centering} 
    \caption{PC5 eigenfunction at y domain}
    \label{subfig:pc5Y}
  \end{subfigure}

  \vspace{0.8em}

    \begin{subfigure}[b]{0.48\textwidth}
    \centering
    \includegraphics[width=\textwidth]{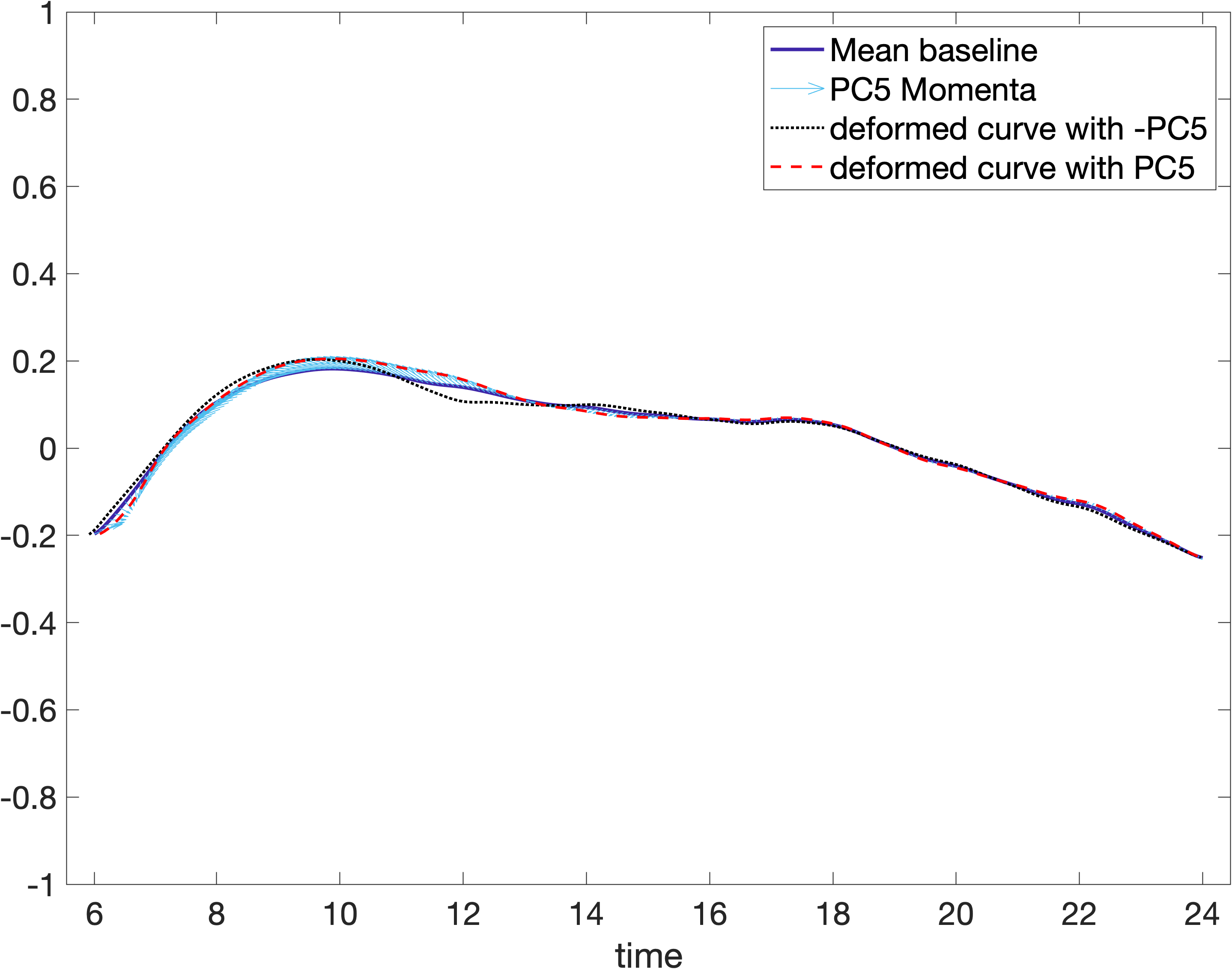}
    \captionsetup{justification=centering} 
    \caption{PC5 eigenfunction based initial momenta and deformations baseline--W1}
    \label{subfig:pc5_V0V1}
  \end{subfigure}
  \begin{subfigure}[b]{0.48\textwidth}
    \centering
    \includegraphics[width=\textwidth]{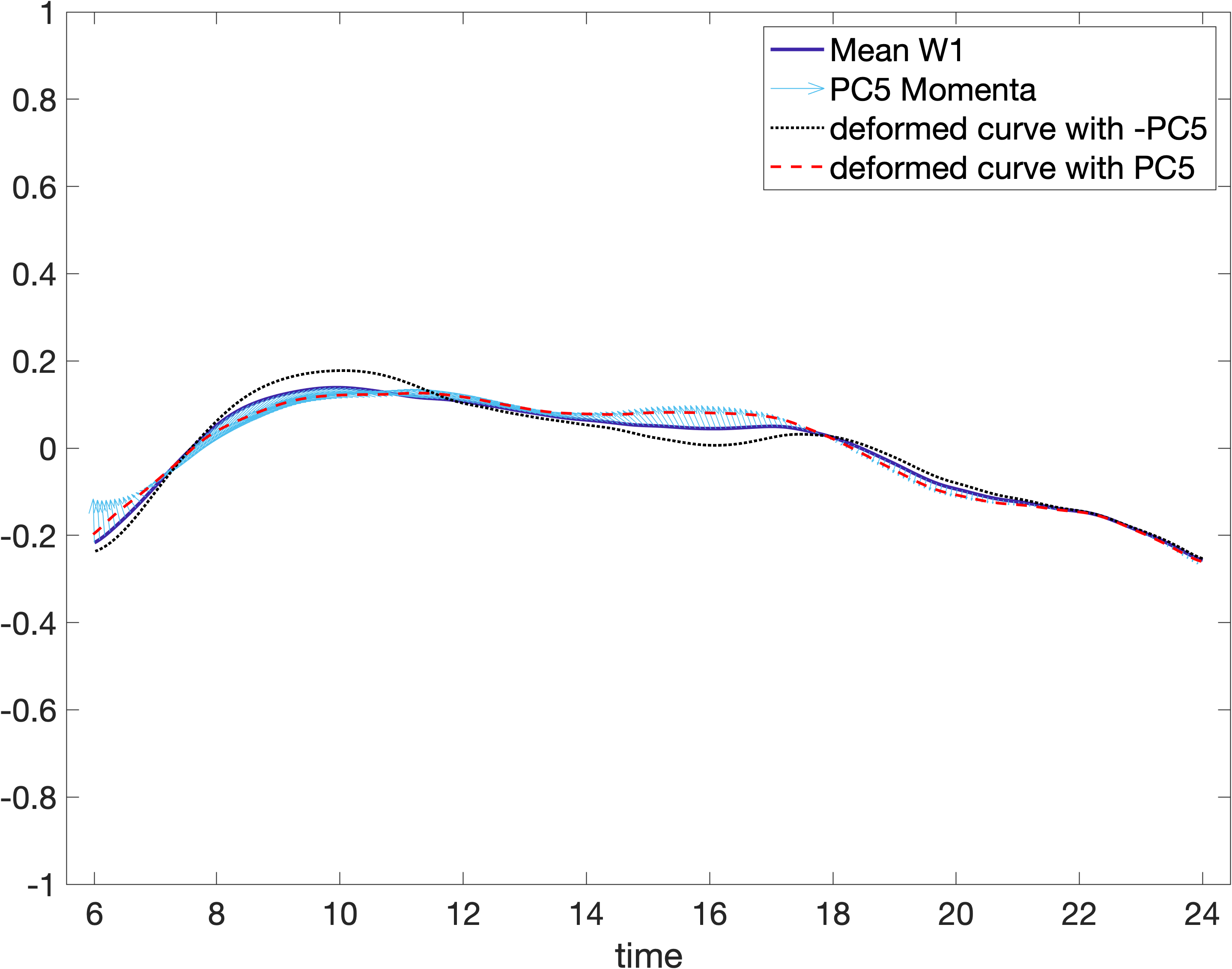}
    \captionsetup{justification=centering} 
    \caption{PC5 eigenfunction based initial momenta and deformations W1--W2}
    \label{subfig:pc5_V1V2}
  \end{subfigure}
    \caption{MFPCA PC5 eigenfunctions and deformations. PC5 shows a temporal left shift during most of the day in both periods. Change of activity levels is very small: there is a slight increase in baseline -- W1 around 9 a.m. to 1:30 p.m. while the increase appears around 12 p.m. to 6 p.m. in W1 -- W2.}\label{fig:MFPCA_efunc5}
\end{figure}

\clearpage
\subsection{Comparison of MFPCA versus Concatenated UFPCA}\label{MFPCAvsUFPCA}

\begin{figure}[h]
  \centering
 \begin{subfigure}[b]{0.47\textwidth}
    \centering
    \includegraphics[width=\textwidth]{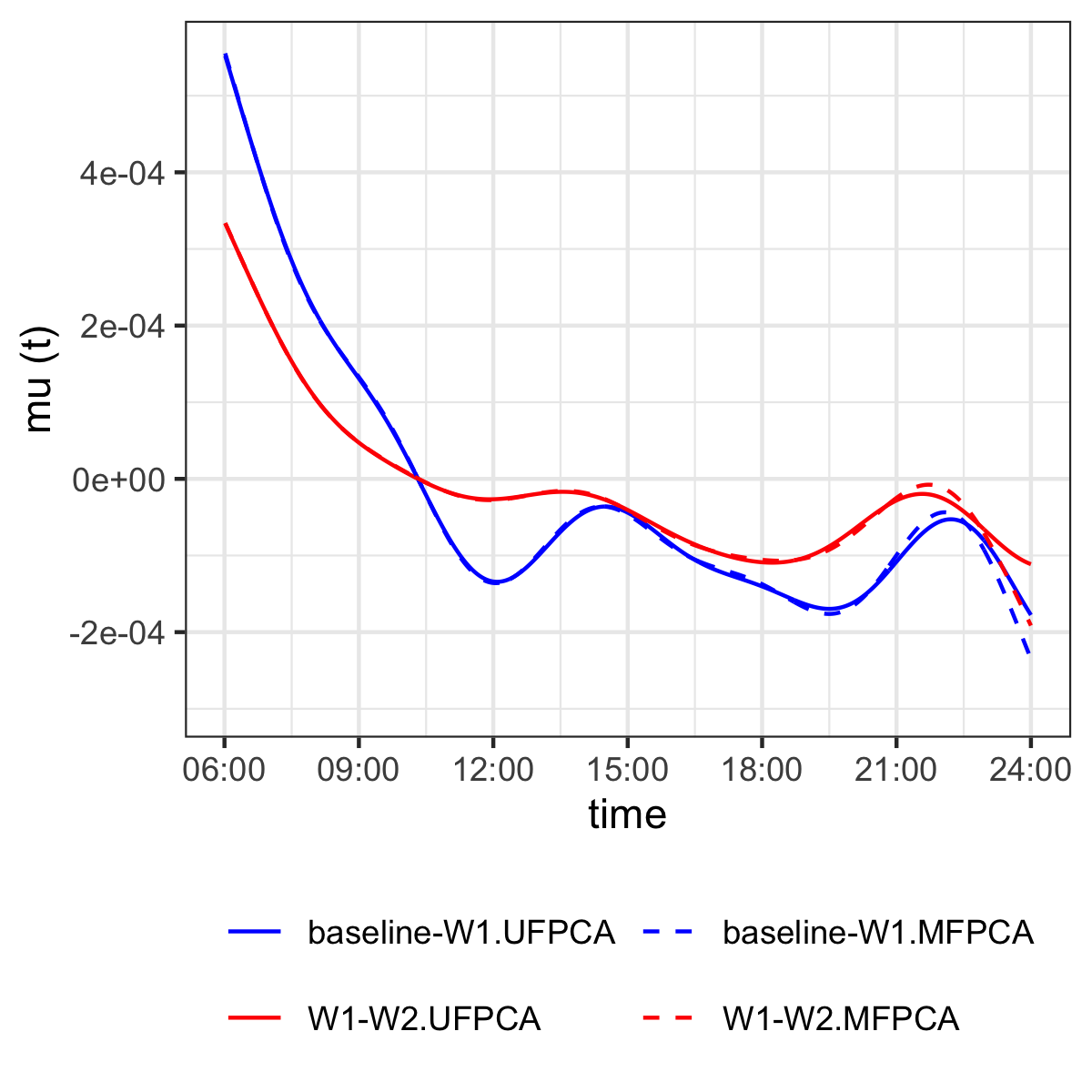}
    \captionsetup{justification=centering} 
    \caption{Mean functions at x domain for both models}
  \end{subfigure}
  \hspace{0.02\textwidth} 
  \begin{subfigure}[b]{0.47\textwidth}
    \centering
    \includegraphics[width=\textwidth]{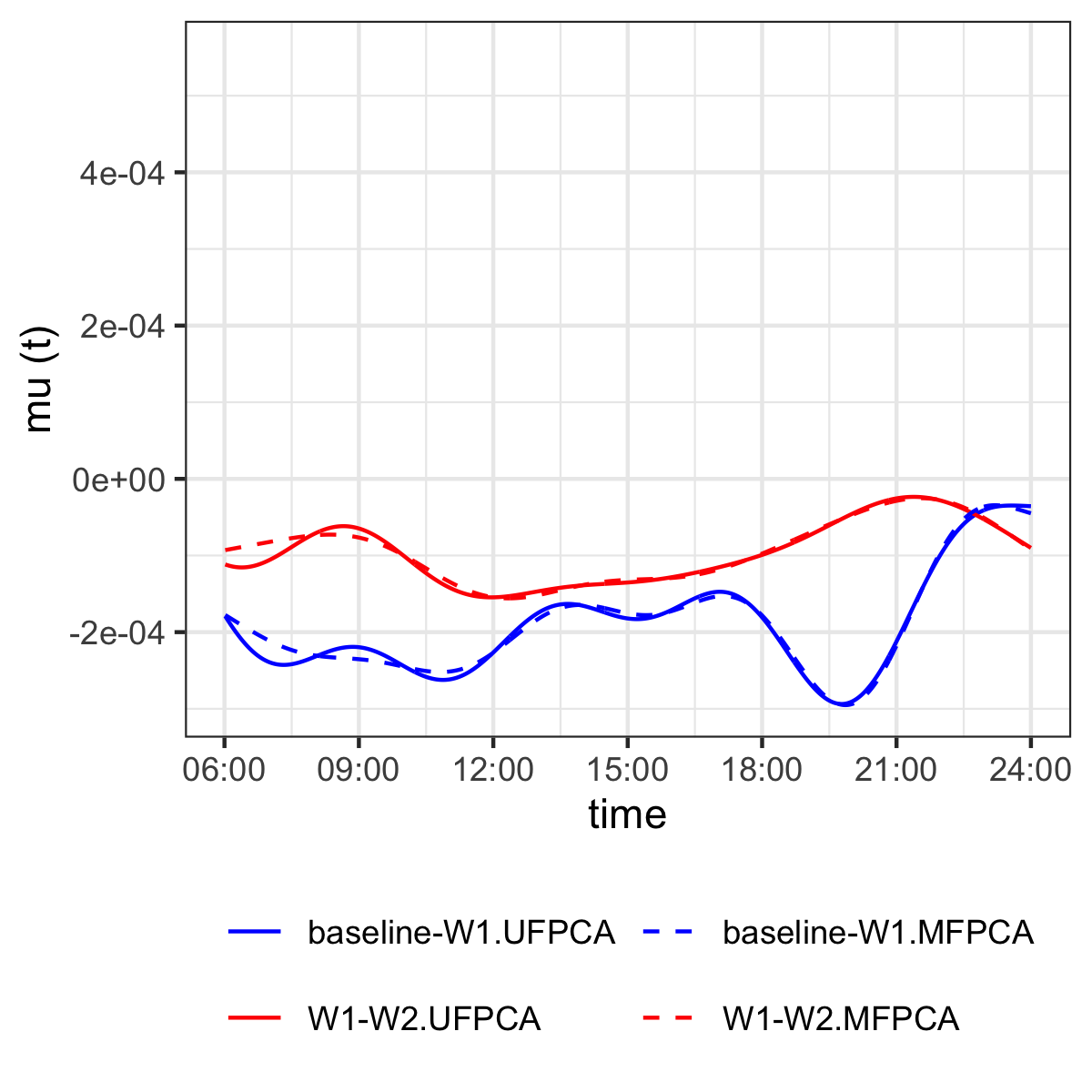}
    \captionsetup{justification=centering} 
      \caption{Mean functions at y domain for both models}
  \end{subfigure}
  
  \vspace{1em}
  
   \begin{subfigure}[b]{0.47\textwidth}
    \centering
    \includegraphics[width=\textwidth]{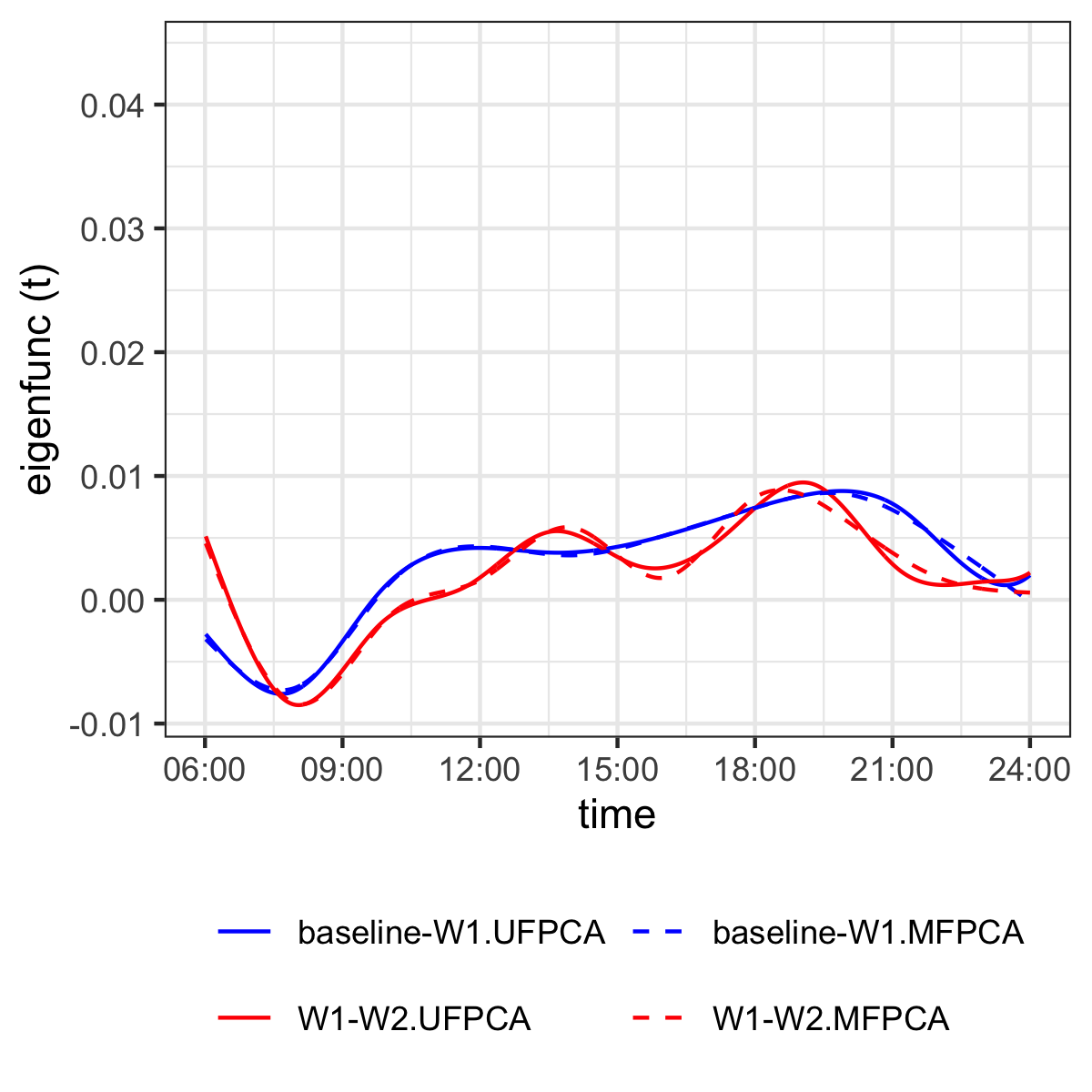}
    \captionsetup{justification=centering} 
    \caption{PC 1 at x domain for both models}
  \end{subfigure}
  \hspace{0.02\textwidth} 
  \begin{subfigure}[b]{0.47\textwidth}
    \centering
    \includegraphics[width=\textwidth]{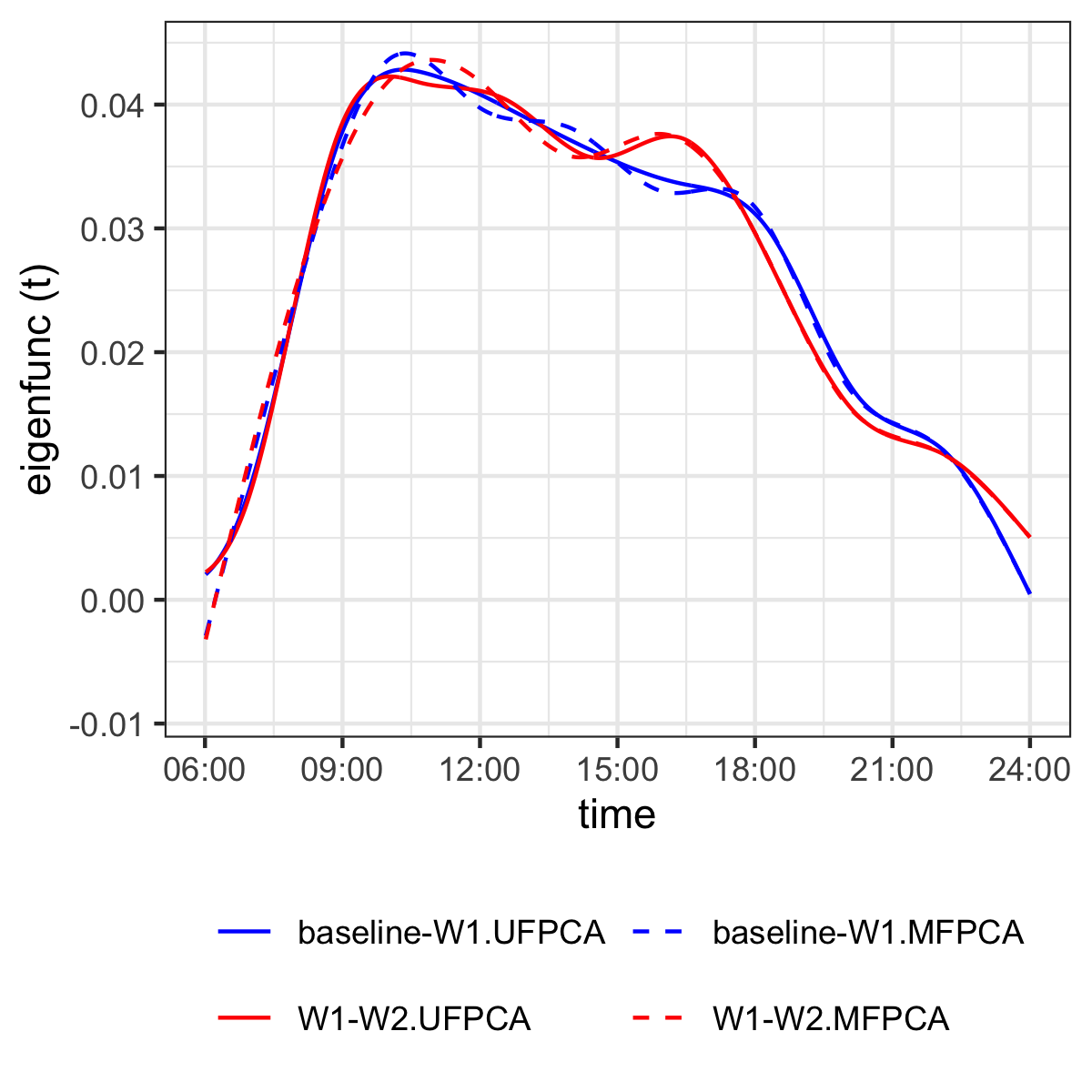}
    \captionsetup{justification=centering} 
      \caption{PC 1 at y domain for both models}
  \end{subfigure}
  \caption{Comparison of MFPCA versus Concatenated UFPCA (PC1)}\label{fig:MFPCAvsUFPCA}

\end{figure}

\begin{figure}[h]
  \centering
 \begin{subfigure}[b]{0.47\textwidth}
    \centering
    \includegraphics[width=\textwidth]{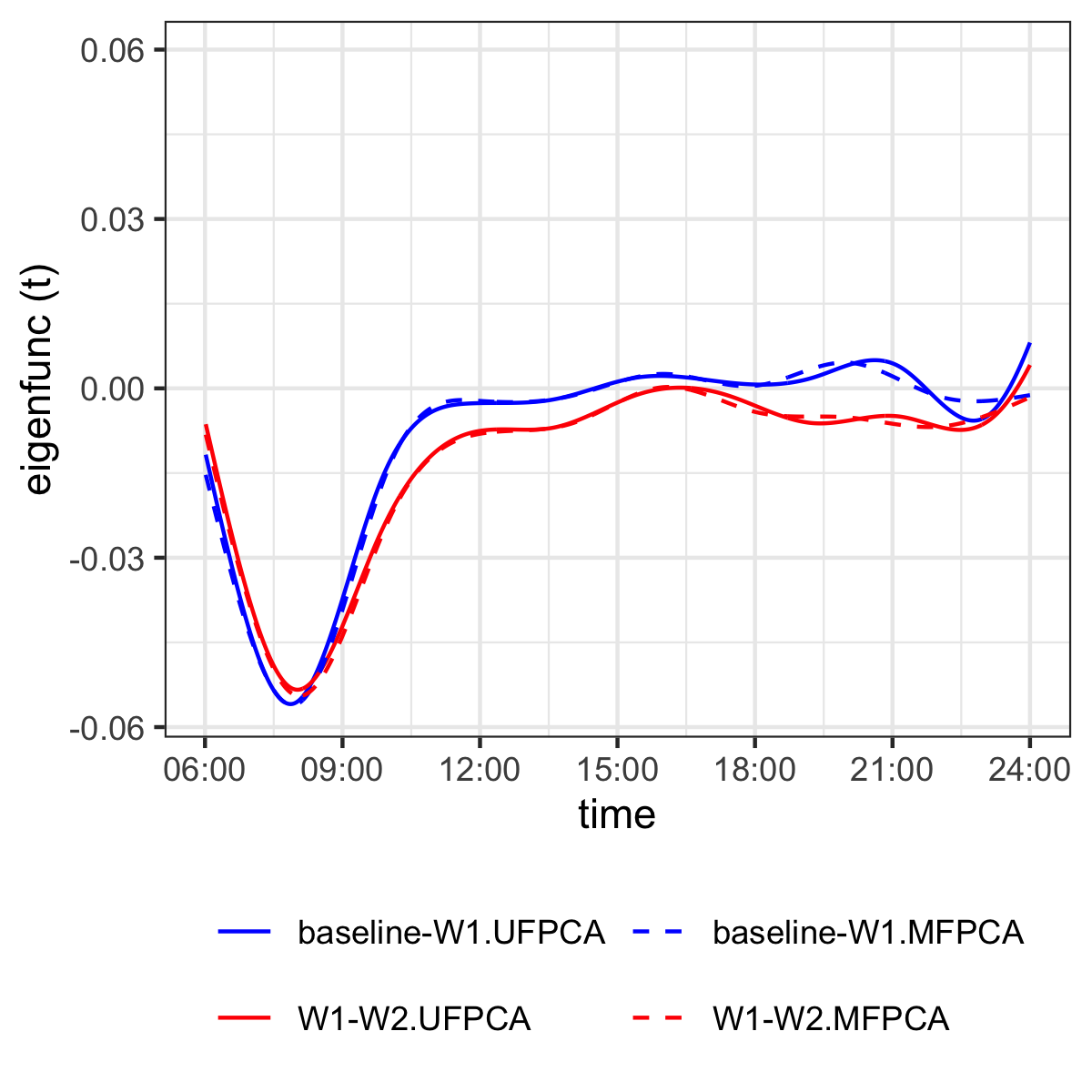}
   \caption{PC 2 at x domain for both models}
  \end{subfigure}
  \hspace{0.02\textwidth} 
  \begin{subfigure}[b]{0.47\textwidth}
    \centering
    \includegraphics[width=\textwidth]{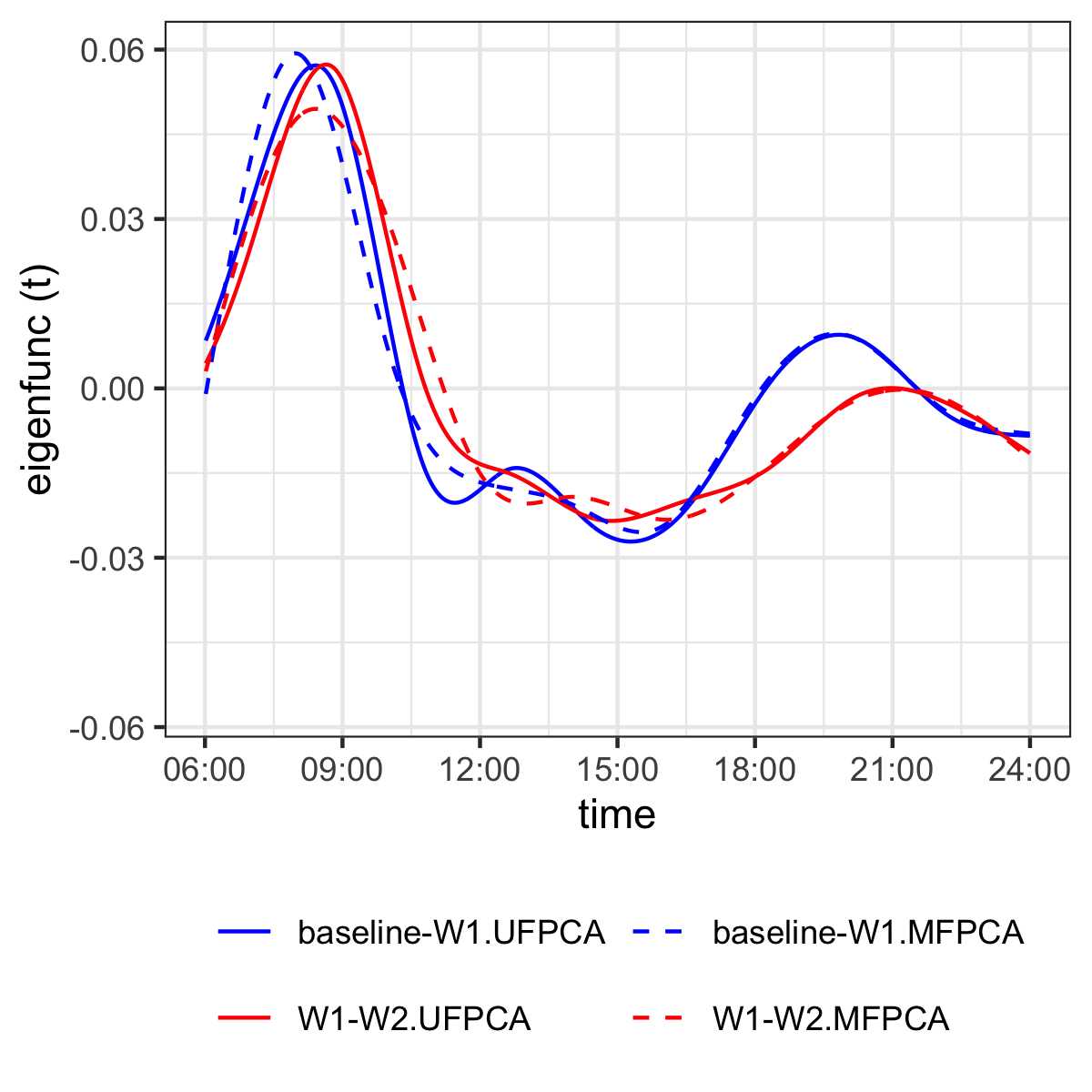}
     \caption{PC 2 at y domain for both models}
  \end{subfigure}
  
  \vspace{0.8em}
  
   \begin{subfigure}[b]{0.47\textwidth}
    \centering
    \includegraphics[width=\textwidth]{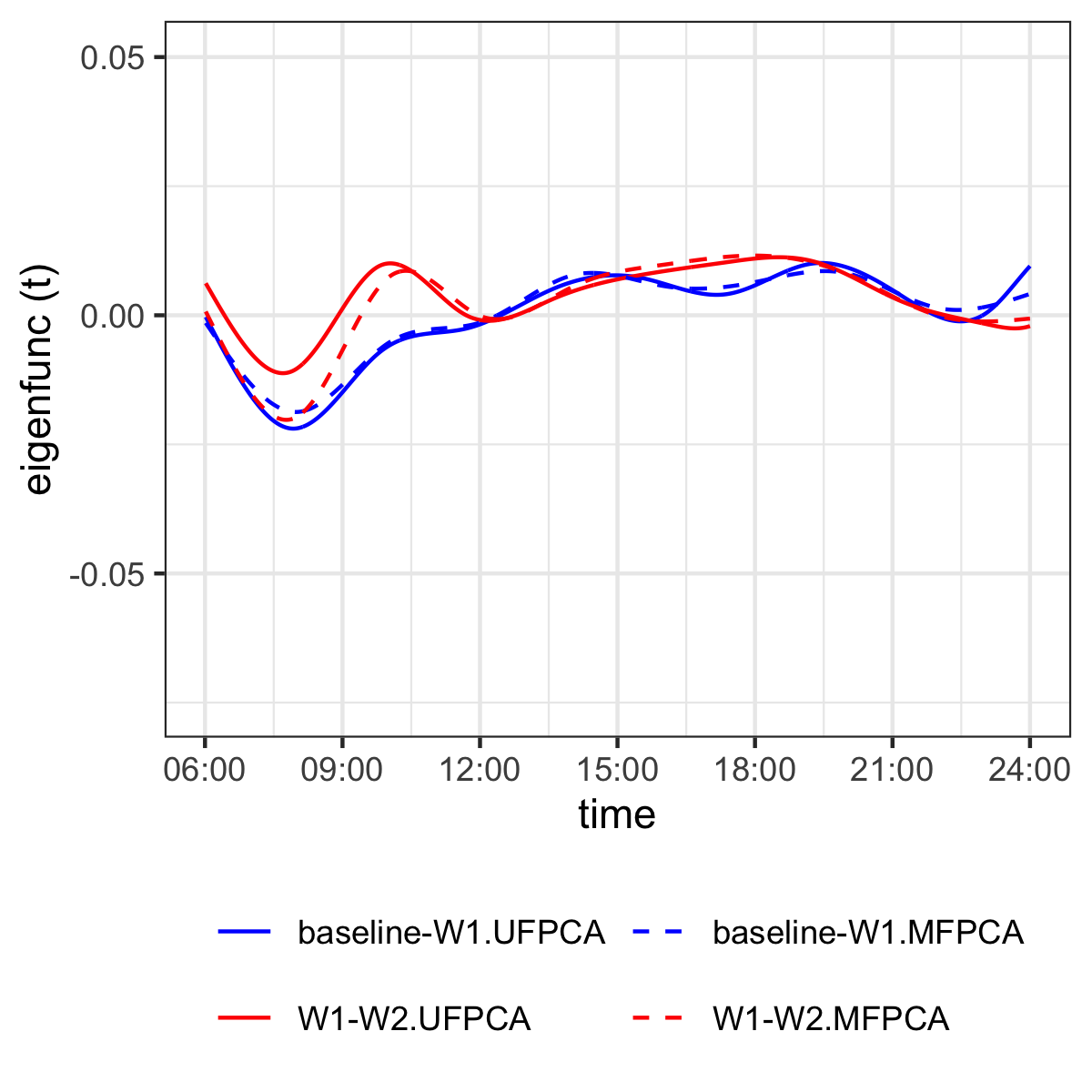}
    \caption{PC 3 at x domain for both models}
  \end{subfigure}
  \hspace{0.02\textwidth} 
  \begin{subfigure}[b]{0.47\textwidth}
    \centering
    \includegraphics[width=\textwidth]{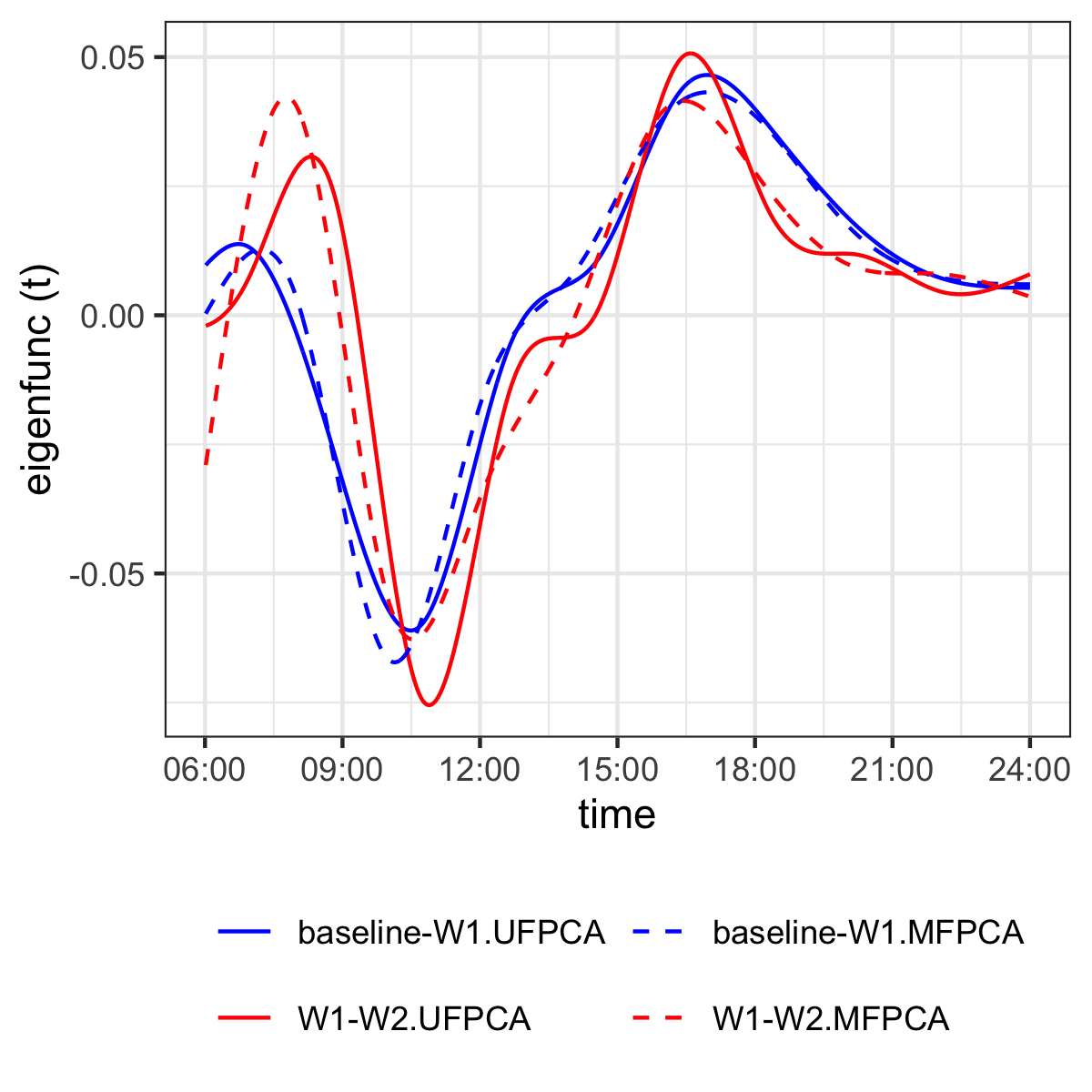}
      \caption{PC 3 at y domain for both models}
  \end{subfigure}
\end{figure}

\begin{figure}[h]\ContinuedFloat
  \centering
   \begin{subfigure}[b]{0.47\textwidth}
    \centering
    \includegraphics[width=\textwidth]{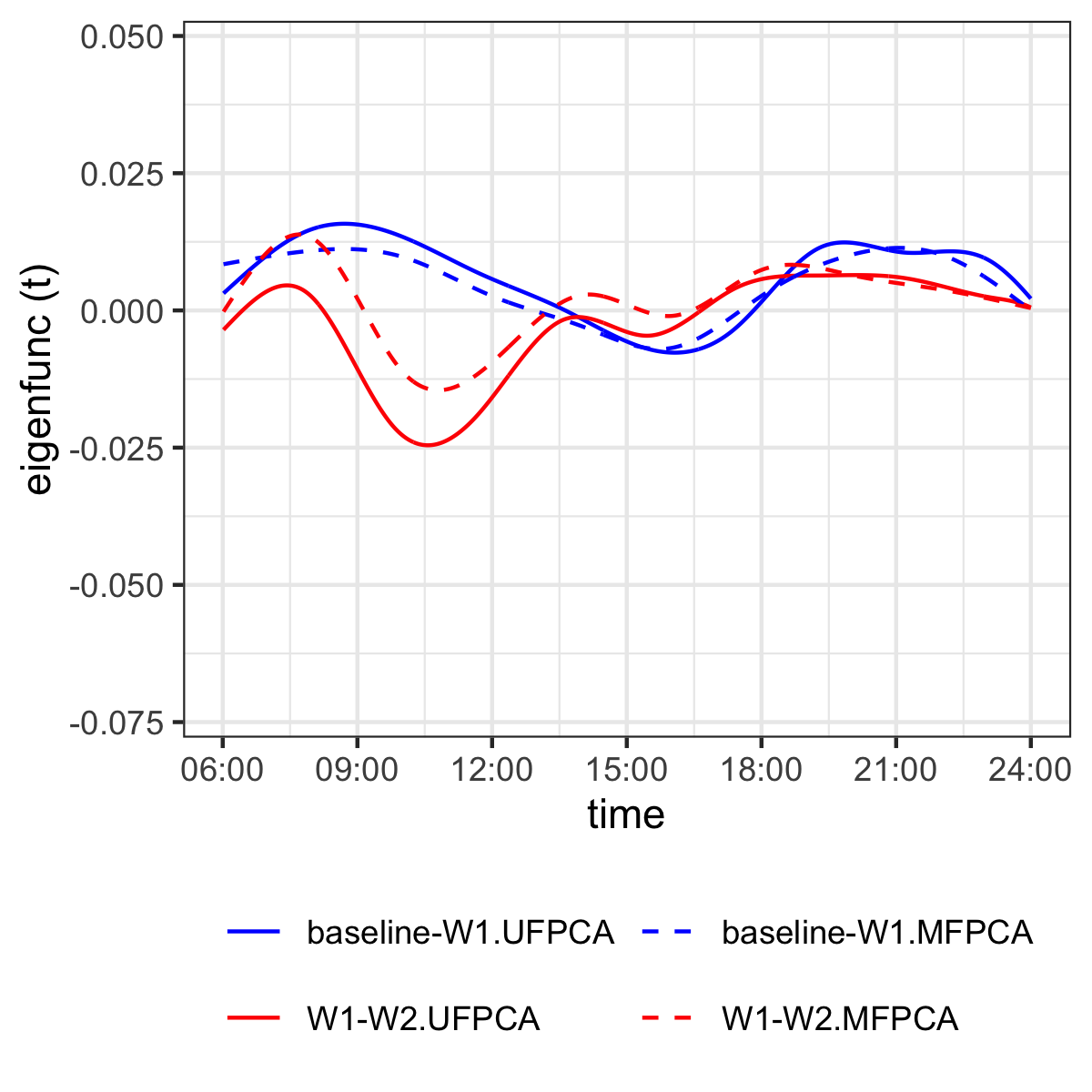}
    \caption{PC 4 at x domain for both models}
  \end{subfigure}
  \hspace{0.02\textwidth} 
  \begin{subfigure}[b]{0.47\textwidth}
    \centering
    \includegraphics[width=\textwidth]{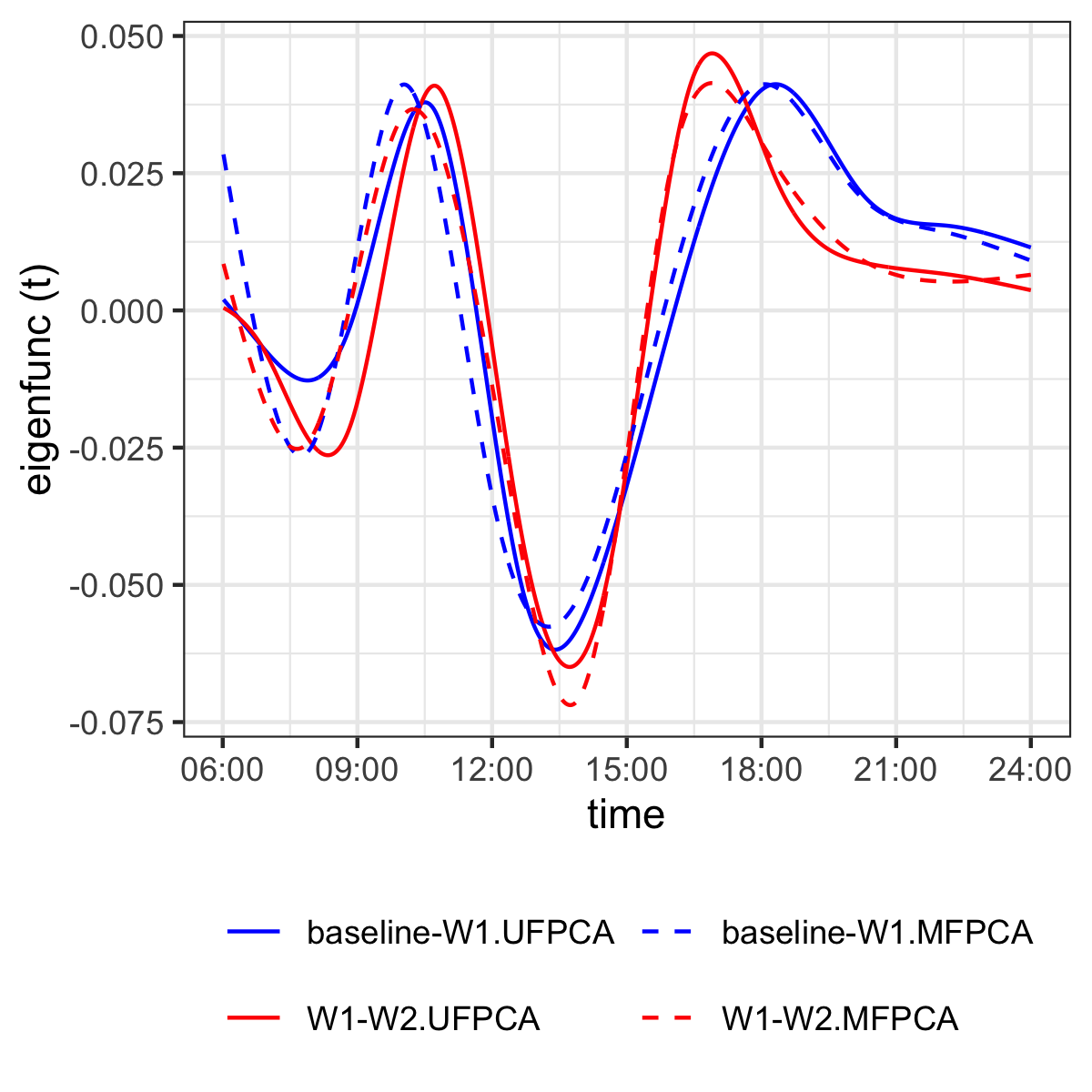}
      \caption{PC 4 at y domain for both models}
  \end{subfigure}
  
   \vspace{0.8em}
   
    \begin{subfigure}[b]{0.47\textwidth}
    \centering
    \includegraphics[width=\textwidth]{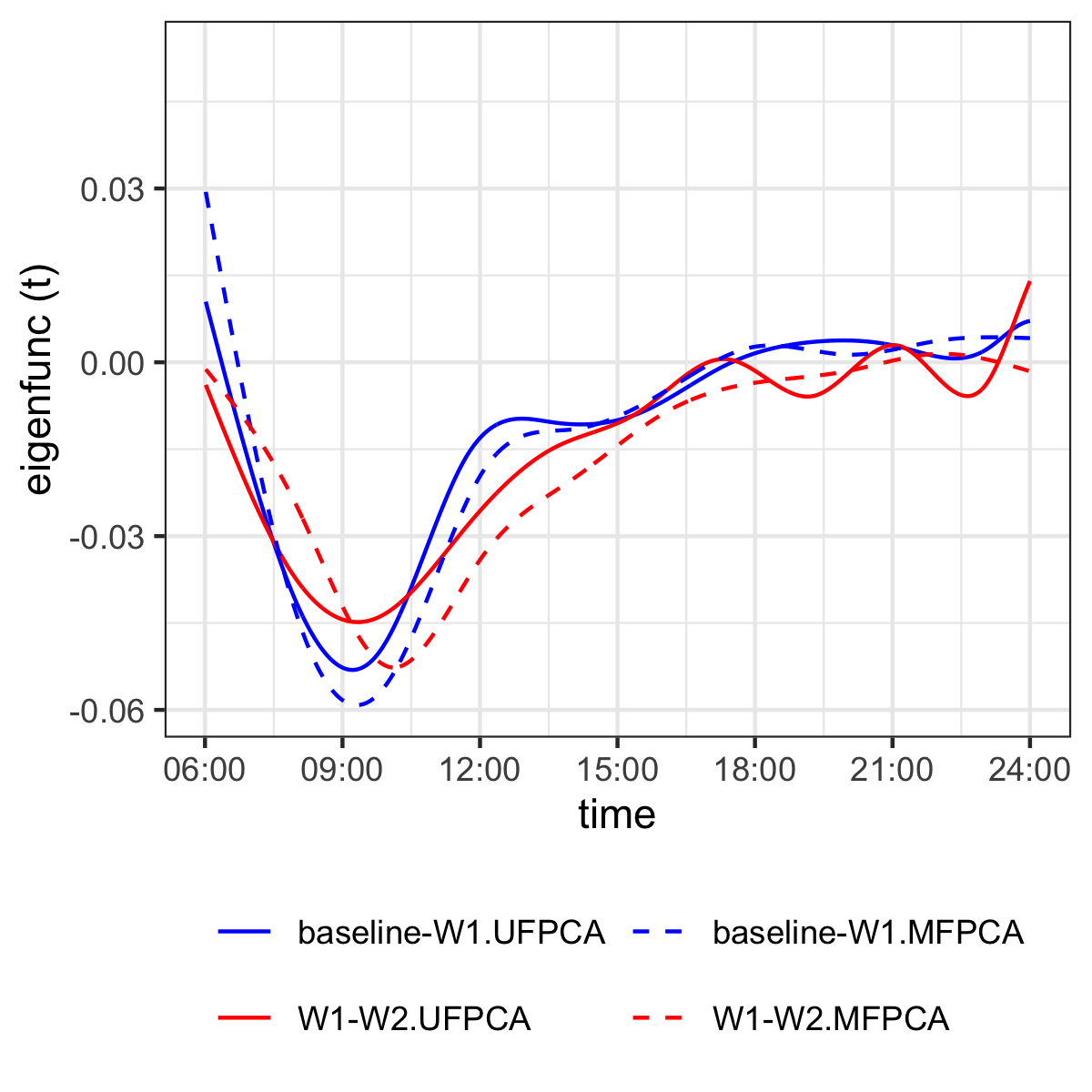}
    \caption{PC 5 at x domain for both models}
  \end{subfigure}
  \hspace{0.02\textwidth} 
  \begin{subfigure}[b]{0.47\textwidth}
    \centering
    \includegraphics[width=\textwidth]{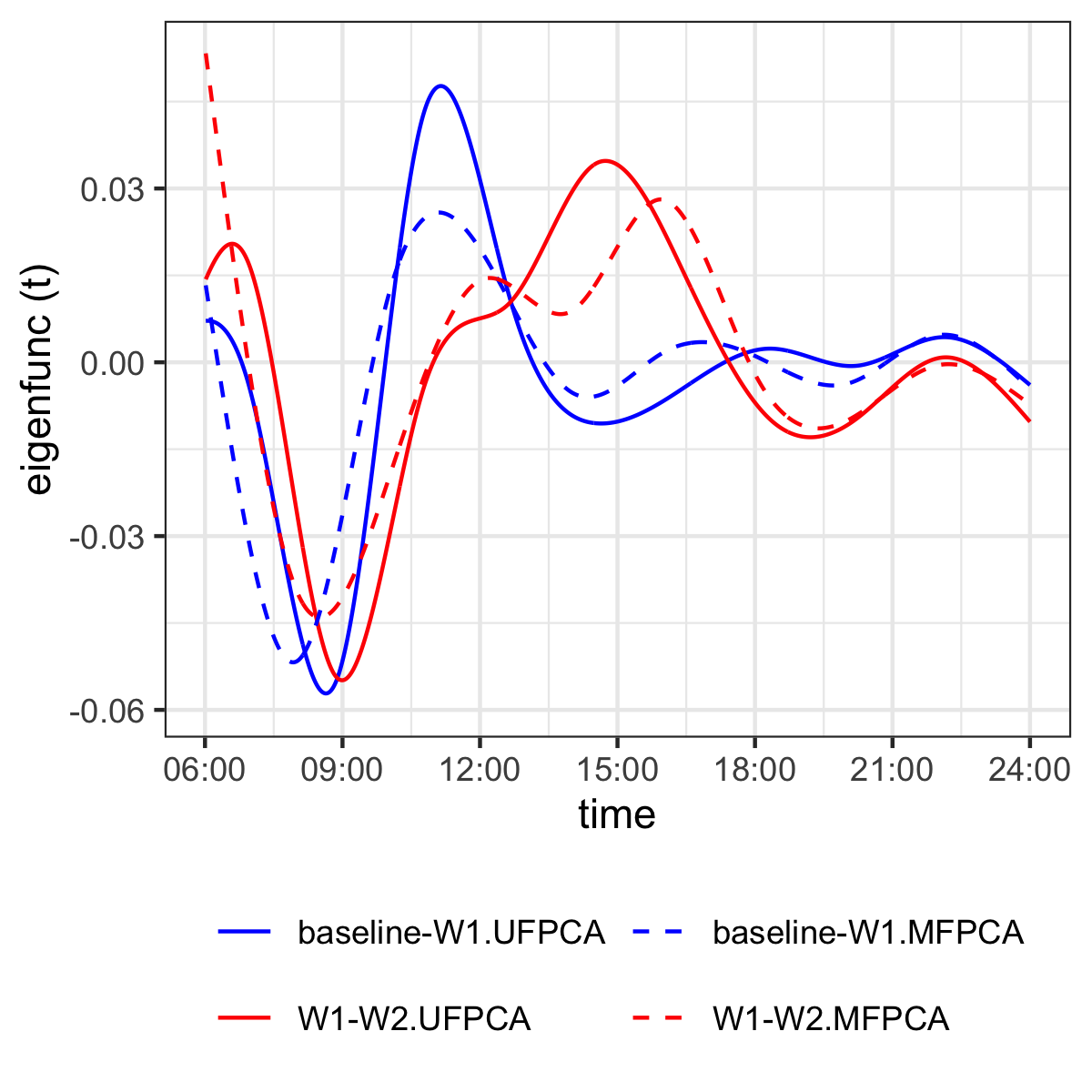}
      \caption{PC 5 at y domain for both models}
  \end{subfigure}
  
  \caption{Comparison of MFPCA versus Concatenated UFPCA (PC2 to PC5)}\label{fig:mfpca_ufpca_pc}

\end{figure}

\clearpage

\end{document}